\documentclass[12pt]{article}
\usepackage{amssymb,amsfonts,amsmath}
\usepackage{latexsym}
\usepackage{mathrsfs}
\usepackage{amssymb,amsfonts,amsmath}
\usepackage{latexsym}
\usepackage{mathrsfs}

\newcommand{\be}{\begin{equation}}
\newcommand{\ee}{\end{equation}}
\newcommand{\bea}{\begin{eqnarray}}
\newcommand{\eea}{\end{eqnarray}}
\newcommand{\beas}{\begin{eqnarray*}}
\newcommand{\eeas}{\end{eqnarray*}}
\newcommand{\pa}{\partial}
\newcommand{\parsh}[2]{\frac{\pa #1}{\pa #2}}

\newcommand{\la}{\lambda}
\newcommand{\qed}{\begin{flushright} $\square$ \end{flushright} }

\newcommand{\ra}{\rightarrow}

\newcommand{\e}{\mathbf{e}}

\usepackage{psfrag}

\usepackage{epsfig,color}
\usepackage{graphicx}

\begin{document}

\title{The Eccentric Frame Decomposition  of Central Force Fields}
\author{Jared M. Maruskin\footnote{Ph.D. Candidate, Department of Mathematics, The University of Michigan}, \ 
Daniel J. Scheeres\footnote{Associate Professor, Department of Aerospace Engineering, The University of Michigan}, \\
Fred C. Adams\footnote{Professor, Department of Physics, The University of Michigan}, \ and 
Anthony M. Bloch\footnote{Professor, Department of Mathematics, The University of Michigan}
}
\maketitle

\begin{abstract} The rosette-shaped motion of a particle in a central force field is known to be classically solvable
by quadratures.  We present a new approach of describing and characterizing such motion based on the eccentricity vector
of the two body problem.  In general, this vector is 
not an integral of motion.  However, the orbital motion, when viewed from the nonuniformly rotating frame defined by the orientation of the 
eccentricity vector, can be solved analytically and will either be a closed periodic circulation or libration.  The motion
with respect to inertial space is then given by integrating the argument of periapsis with respect to time.  Finally we will 
apply the decomposition to a modern central potential, the spherical Hernquist-Newton potential, which models dark matter
halos of galaxies with central black holes.
\end{abstract}

\section{Introduction}

\subsection{Central Force Fields}

The motion of a particle in a central force field is known to be classically solvable by quadratures.  Due to the 
spherical symmetry of the force field, an angular momentum integral exists and the ensuing motion is
confined to a single orbital plane so that, without loss of generality, we can assume the system to have
 2 degrees of freedom.  In polar coordinates, the Hamiltonian may be expressed as:
\[
H = \frac 1 2 \left( \dot r^2 + \frac{h^2}{r^2} \right) - U(r),
\]
where  $h = r^2 \dot \theta$ is the angular
momentum and $U(r)$ is the potential energy function.  We  will use the standard convention that dots refer to 
time derivatives, whereas primes refer to spatial derivatives.  The corresponding Hamiltonian system,
\[
\dot r = v_r  \qquad  \dot v_r = \frac{h^2}{r^3} + U'(r) \qquad \dot \theta = \frac{h}{r^2} \qquad \dot h = 0,
\]
is integrable by quadratures:
\be
\label{ecc27}
\int_{r(0)}^{r(t)} \frac{ \pm dr }{\sqrt{2H + 2U(r) - h^2/r^2} } = t - t_0
\ee
\be
\label{ecc28}
\theta(t) = \theta_0 + \int_{t_0}^t \frac{h \ dt}{r(t)^2}.
\ee
The ensuing motion follows rosette-shaped paths (Arnold (1989), Whittaker (1988), etc.).

\subsection{Osculating Orbital Elements}

One could, alternatively, proceed using Variation of Parameters and Lagrange's Planetary Equations 
(Brouwer \& Clemence (1961), Roy 1988).  In this case, one can write down differential
equations of motion for the six osculating classical orbital elements and then solve them by quadrature for all time.  
These equations are nonlinear and furthermore depend upon a choice of the 
``planetary'' gravitational parameter $\mu$.  One assumes the 
potential is a perturbation of a Newtonian potential:
\[
U(r) = \frac{\mu}{r} + R(r),
\]
where $R(r)$ is known as the \textit{disturbing function}.  For a general central force field where there
is no nominal attracting body, such a choice is somewhat arbitrary.  For instance, there is 
no primary ``planet'' when considering motion in a galactic halo.  A gravitational parameter for the unperturbed
motion can nonetheless be artificially contrived, perhaps based on the total halo mass (if the motion evolves
in the outskirts of the galaxy) or based on the mass of a central galactic bulge or black hole (if the motion
evolves near the galactic core).  Whatever the choice of gravitational parameter, a complete set of osculating
orbital elements arises and the ensuing motion can be determined.

\subsection{The Eccentric Frame}

We will define a gravitational parameter based on the central force field's potential, with no reference
to a main attracting body and perturbation theory.  Following the classical analogy, we define $\mu(r)$ such that:
\[
U(r) = \frac{\mu(r)}{r}.
\]
We will show that this gives rise to a nonstatic eccentricity vector that rotates at a nonuniform rate.
The eccentricity vector (Runge-Lenz vector) associated with this spatially variable gravitational parameter function defines
a preferred coordinate system which we call the eccentric frame.
With respect to this frame, we will show that the motion follows a closed orbit.  Depending on the value of energy,
the particle will make closed circulations or librations in the eccentric frame.  The eccentric frame decomposition
gives rise to a set of orbital elements.  We will discuss their physical implications and the key features of how they arise.
In particular, one can have circular orbits in inertial space with nonzero osculating eccentricity.  This feature is not 
unique to our method, it can arise from any choice of osculating orbital elements.  The eccentric frame decomposition,
however, illuminates the behavior and gives rise to a new standard description that better fits orbits of central force
field potentials.


\section{The Eccentric Frame Decomposition}

We first define the eccentric frame by means of specifying the nonstatic eccentricity vector 
associated with the gravitational parameter function $\mu(r) = r U(r)$.  We then show that the particle
traces a closed orbit as viewed from this noninertial frame.  Finally we compute the set of 
osculating orbital elements that belong to this system.

\subsection{Motion with respect to the Eccentric Frame}

Given a spherically symmetric potential energy field, we can recast the Hamiltonian into
the following form, reminiscent of its classical analogy:
\be
\label{ecc01}
E = \frac 1 2 \left( v^2 + \frac{h^2}{r^2} \right) - \frac{\mu(r)}{r},
\ee
where $h= r^2 \dot \theta$ is the magnitude of the angular momentum vector, 
\be
\label{ecc02}
\mathbf{H} = \mathbf{r} \times \dot{\mathbf{r}},
\ee
and $(r,v, \theta, h)$ are
the symplectic coordinates, with $v = \dot r$.  This gives rise to the following Hamiltonian equations of motion:
\beas
\dot r = v & \qquad & \dot v = \frac{h^2}{r^3} + \frac{\mu'(r)}{r} - \frac{\mu(r)}{r^2} \\
\dot \theta = \frac{h}{r^2} & \qquad & \dot h = 0
\eeas
which can be recast in the following form
\be
\label{ecc03}
\ddot{\mathbf{r}} = \left( \ddot r - \frac{h^2}{r^3} \right) \e_r = \left(
\frac{\mu'(r)}{r} - \frac{\mu(r)}{r^2} \right) \e_r,
\ee
where $\mathbf{r} = r \e_r$.  
Consider now the eccentricity vector
\be
\label{ecc04}
\mathbf{B} = \dot{\mathbf{r}} \times \mathbf{H} - \mu(r) \e_r.
\ee
Its evolution is governed by the following equations of motion:
\beas
\dot{\mathbf{B}} &=& \ddot{\mathbf{r}} \times \mathbf{H} - \mu'(r) \dot r \e_r - \mu(r) \dot \theta
\e_\theta \\
&=& \left( \frac{\mu'(r)}{r} - \frac{\mu(r)}{r^2} \right) r^2 \dot \theta \e_r \times
\hat{\mathbf{H}} - \mu'(r) \dot r \e_r - \mu(r) \dot \theta
\e_\theta \\
&=& - \left( \frac{\mu'(r)}{r} - \frac{\mu(r)}{r^2} \right) r^2 \dot \theta \e_\theta
 - \mu'(r) \dot r \e_r - \mu(r) \dot \theta
\e_\theta \\
&=& - \mu'(r) \dot{\mathbf{r}}.
\eeas
The vector $\mathbf{B}$ itself works out to be
\beas
\mathbf{B} &=& r^2 \dot \theta (\dot r \e_r + r \dot \theta \e_\theta ) \times
\hat{\mathbf{H}} - \mu(r) \e_r \\
&=& - r^2 \dot r \dot \theta \e_\theta + r^3 \dot \theta^2 \e_r - \mu(r) \e_r \\
&=& -h \dot r \e_\theta + \left( \frac{h^2}{r} - \mu(r) \right) \e_r.
\eeas
We thus find the magnitude of $\mathbf{B}$ is:
\be
\label{ecc05}
B = \sqrt{2h^2E + \mu^2(r)}.
\ee
We define the argument of periapsis, $\omega$, to be the angle made between the inertial $x$-axis and the 
$\mathbf{B}$-vector.  The $\mathbf{B}$-vector defines a rotating reference frame, which we call the eccentric frame.
We define $\hat{\mathbf{B}}$ to be a unit vector in the $\mathbf{B}$ direction.  Hats will denote unit vectors.  
Let $X$ and $Y$ be the cartesian coordinates of the particle with respect to the eccentric frame and let $x$ and $y$
be the cartesian coordinates of the particle with respect to the inertial frame.  The axes of the inertial
frame are determined by the stationary
unit vectors $\hat{\mathbf{i}}$ and $\hat{\mathbf{j}}$.
The polar angle of the particle measured with respect to the $\hat{\mathbf{B}}$ direction is known as the 
true anomaly $f$.  This notation is also used in Roy (1988).
The polar angle of the particle in the inertial frame is
related to the true anomaly by the following relation:
\be
\label{ecc06}
\theta = f + \omega.
\ee
In some of the literature, the true anomaly $f$ is denoted by $\nu$; and the inertial polar angle (argument of
latitude) $\theta$ is denoted by $u$. 
Decomposing the eccentricity vector $\mathbf{B}$ in the inertial frame, we see that
\beas
\mathbf{B} &=& \left( \dot r h \sin \theta + \left( \frac{h^2}{r} - \mu(r) \right) \cos \theta \right) \mathbf{\hat{i}} \\
& & + \left( - \dot r h \cos \theta + \left( \frac{h^2}{r} - \mu(r) \right) \sin \theta \right) \mathbf{\hat{j}}.
\eeas
However, by definition, $\mathbf{B} = B(r) (\cos \omega \hat{\mathbf{i}} + \sin \omega \hat{\mathbf{j}})$.  Hence:
\bea
\left[ \begin{array}{c}
\cos \omega \\ \sin \omega \end{array} \right] &=&
\frac{1}{B(r)} \mathbf{A} 
\cdot
\left[ \begin{array}{c}
\cos \theta \\
\sin \theta \end{array} \right] \nonumber \\
\label{ecc07}
& =&
 \frac{1}{B(r)} \mathbf{A} 
\cdot
\mathbf{F} \cdot
\left[ \begin{array}{c}
\cos \omega \\
\sin \omega \end{array} \right]
\eea
where we have defined
\[
\mathbf{A} = 
 \left[ \begin{array}{cc}
\left( h^2/r - \mu(r) \right) & \dot r h \\
- \dot r h & \left( h^2/r - \mu(r) \right) \end{array} \right]
\]
and
\[
\mathbf{F} = \left[ \begin{array}{cc}
\cos f & - \sin f \\
\sin f & \cos f \end{array}
\right]
\]
and have further made use of the trigonometric identities
\beas
\cos \theta &=& \cos( f + \omega) = \cos f \cos \omega - \sin f \sin \omega \\
\sin \theta &=& \sin(f + \omega) = \sin f \cos \omega + \cos f \sin \omega.
\eeas
We recognize that the matrix premultiplying the vector $\langle \cos \omega, \sin \omega \rangle$ on the 
right hand side of \eqref{ecc07} must be the identity matrix.  Hence we have found an explicit expression
relating the true anomaly and the radius:
\bea
\label{ecc08}
\cos f &=& \frac{1}{B(r)} \left( \frac{h^2}{r} - \mu(r) \right) \\
\label{ecc09}
\sin f &=& \frac{1}{B(r)} \dot r h.
\eea
Thus we see that the particle traces out a closed path in the eccentric frame.  
By carefully considering \eqref{ecc09}, we see that periapsis is always achieved at $f = 0$, i.e. when
$\mathbf{r}$ and $\mathbf{B}$ are parallel; and that apoapsis is achieved at $f = \pi$, i.e.
when $\mathbf{r}$ and $\mathbf{B}$ are anti-parallel.

If the angular momentum is positive, \eqref{ecc09} tells us that $r$ is increasing when the particle
is in the upper half plane and is decreasing when the particle is in the lower half plane.  The opposite
is true for the case of a negative angular momentum.

\subsection{The Osculating Eccentricity and Semi-Major Axis}

We would also like to point out that one can rearrange \eqref{ecc08} into the following form:
\be
\label{ecc22}
r = \frac{h^2/\mu(r)}{1 + B(r) \cos f/\mu(r)} = \frac{p(r)}{1 + e(r) \cos f },
\ee
completely analogous to its classical ($\mu(r) = \mbox{const.}$) form.

Utilizing the relation $p = a(1-e^2)$, we can define the osculating eccentricity
and semi-major axis of the system
in closed form as follows:
\bea
\label{ecc25}
e(r) &=& \frac{B(r)}{\mu(r)} \\
\label{ecc26}
a(r) &=& \frac{h^2 \mu(r)}{\mu(r)^2 - B(r)^2}.
\eea

These are given unambiguously as a function of $r$, without integrating.
They represent a standard decomposition of the motion.  
Using a standard choice of osculating orbital elements,
one would first define a semi-arbitrary choice for a fixed $\mu$.  Thus, there is no unique
standard set of osculating orbital elements for a general system, rather a one parameter family
of orbital elements that describe the motion.  By using the radially varying $\mu(r)$,
we seek to better normalize the description of motion in such systems.

Since the true anomaly is given by the relations \eqref{ecc08} and \eqref{ecc09},
one now only need solve for the osculating argument of periapsis to obtain the complete
motion as a function of time.

\subsection{The Osculating Argument of Periapsis}

Solving for the osculating argument of periapsis can be done in one of two ways.  First, one may integrate
\eqref{ecc27}-\eqref{ecc28} by quadratures.  Once $r$ and $\theta$ are known, $f$, $a$, and $e$
can be extracted by the above relations \eqref{ecc08}, \eqref{ecc09}, \eqref{ecc25}, \eqref{ecc26};
then the osculating argument of periapsis can be solved by means of the relations
$\theta = f + \omega$.  On the other hand one can solve the quadrature we derive below.

To determine the rotation of the eccentric frame, consider the angular momentum integral:
\be
\label{ecc10}
h = r^2 \dot \theta = r^2 (\dot f + \dot \omega) = r^2 f'(r) \dot r \left( 1 + \frac{d\omega}{df} \right).
\ee
Differentiating \eqref{ecc08} and utilizing \eqref{ecc09}, we have that
\beas
- \sin(f(r)) f'(r) &=& \frac{-\dot r f'(r) h}{B(r)} \\
&=& \frac{1}{B(r)} \left( \frac{-h^2}{r^2} - \mu'(r) \right) \\
& & - \frac{B'(r)}{B(r)^2} \left( \frac{h^2}{r} - \mu(r) \right).
\eeas
We can now solve \eqref{ecc10} for $\omega'(f)$:
\be
\label{ecc11}
\frac{d\omega}{df} = \frac{h}{f'(r) \dot r r^2} - 1
= \frac{\Phi(r)}{B(r)h^2 - \Phi(r)}
\ee
where we define $\Phi(r)$ as:
\be
\label{ecc12}
\Phi(r) = \mu(r) r^2 B'(r) - B'(r) r h^2 - \mu'(r) r^2 B(r).
\ee
We thus have
\be
\label{ecc13}
\omega(f) = \omega(0) + \int_0^f \frac{\Phi(r)}{B(r)h^2 - \Phi(r)} \ d\tilde f.
\ee
where we recognize $r = r(\tilde f)$ in the integrand, by the relations \eqref{ecc08} and \eqref{ecc09}.
Together with \eqref{ecc25} and \eqref{ecc26}, this constitutes a full set of osculating orbital elements
that are well-defined for the orbit for all time.

The full motion is then completely specified in terms of the parameter $f$ by the relation:
\[
\theta(f) = f + \omega(f).
\]

\section{The Zero Velocity Curve}

The central force problem is a 2 degree of freedom problem with 2 integrals of motion, $E$ and $h$.
It is therefore integrable and, in fact, reduces to motion on a Liouville torus.  The symplectic
coordinates of the system are $(r, v, \theta, h)$.  The coordinate $h$ is conserved, and the motion therefore takes place
on the $h= \mathrm{const.}$ hyper-plane.  Motion in the $(r,v)$ plane is constrained to the 
curve $\Gamma_{h, E}$ defined by \eqref{ecc01}, with fixed $E$ and $h$.  Meanwhile, $\theta$ cycles along $S^1$
according to $h = r^2 \dot \theta$.  Motion in the reduced $(r,v,\theta)$ space can therefore be visualized
as follows:  it is constrained to the surface obtained by revolving the curve $\Gamma_{h, E}$ around the $v$
axis.  This resulting surface is (obviously) topologically equivalent to the Liouville torus, but obtained 
directly without the arduous task of computing action-angle variables.

For a fixed $h$, as one varies the energy, one encounters various bifurcation points where the system undergoes
changes.

\subsection{Periapsis and Apoapsis}

Computation of the periapsis and apoapsis radii is accomplished by the standard technique of 
plotting the zero-velocity curve on the $E-r$ plane.  The plot is obtained by setting $v = \dot r = 0$ in \eqref{ecc01}.
The resulting equation is:
\be
\label{ecc18}
E_{\rm zv}(r) = \frac 1 2 \frac{h^2}{r^2} - \frac{\mu(r)}{r}.
\ee
For a fixed energy $E$, the solutions to this 
equation represent the periapsis $r_p$ and apoapsis $r_a$ radii.  
Maximum and minimum values of of $E_{\rm zv}(r)$ correspond to unstable and stable circular
orbits, respectively.  
If there are multiple ``wells,''
the corresponding roots of this equation alternate $r_{p1}, r_{a1}, r_{p2}, r_{a2}, \ldots$, and the forbidden
regions of the inertial x-y plane are concentric, circular annuli.

\subsection{Circular Orbits}

As one decreases the energy for a fixed angular momentum, the curves $\Gamma_{h, E}$ on the $(r,v)$-plane
shrink until they degenerate to a single point on the $r$-axis which corresponds to a circular orbit in the 
$(r,\theta)$ polar plane.  This occurs at the local minima of $E$ on the $(r,v)$ plane, and hence is given 
by $\nabla E = 0$, where $E$ is given by \eqref{ecc01} and $h$ is held fixed.  This condition amounts to
\bea
v &=& 0 \\
\label{ecc16}
\frac{h^2}{r^3} + \frac{\mu'(r)}{r} - \frac{\mu(r)}{r^2} &=& 0 = \ddot r.
\eea
The root of \eqref{ecc16}, $r_{\rm circ}$, corresponds to the radius of the circular orbit which 
occurs at the minimum energy $E_{\rm circ} := E_{\rm zv}(r_{\rm circ})$.  

\subsection{Escape Orbits}

If $U(r) \ra \mbox{const.}$ as $r \ra \infty$, a series of unbounded orbits are present in the solution space.
Such orbits are classified as escape orbits.  Typically one takes the potential at infinity to be zero, so that
$U(r) \ra 0$ as $r \ra \infty$, so that orbits with negative energies are gravitationally bounded to the center
of the potential, whereas orbits with positive energies have enough energy to escape to infinity.

\section{Circulations vs. Librations in the Eccentric Frame}

As one decreases the energy from $E_{\rm esc}$ to the minimum energy $E_{\rm circ}$, one encounters
a bifurcation in the eccentric frame at $E_{\rm crit}$, as the orbits (as seen from the eccentric frame) 
change from circulations to  librations.  This is a necessary transition
that must occur, as one lowers the energy, before one can reach a circular orbit.
It will be our goal in this section to understand the how this
bifurcation comes about and to give a qualitative description of motion in the eccentric frame for fixed $h$ as one varies $E$.

\subsection{The Critical Energy}

We now define a critical radius and critical energy.
The critical radius is defined as the root to the right hand side of \eqref{ecc08}, which occurs 
when:
\be
\label{ecc17}
h^2 = r \mu(r).
\ee
For a fixed $h$, let the solution to \eqref{ecc17} be $r_{\rm crit}$.  Further, let us define the critical energy as follows:
\be
E_{\rm crit} = E_{\rm zv}(r_{\rm crit}) = - \frac 1 2 \frac{\mu(r_{\rm crit})}{r_{\rm crit}},
\ee
where $E_{\rm zv}(r)$ is given by \eqref{ecc18}.  In the following subsections, we will see how passing through this value
of energy brings about a bifurcation in our system.

The critical radius $r_{\rm crit}$ has an important physical signifigance in terms of the eccentric frame.
From \eqref{ecc08}, we see that $\cos f = 0$, i.e. the particle is crossing the $Y$-axis in the eccentric
frame, exactly when $r = r_{\rm crit}$.  It is interesting to note that $r_{\rm crit}$ is \textit{independent} of the energy of the system.  
Thus, as one changes the energy, the particle passes between the left and right hand planes through the same two portals
($Y = \pm r_{\rm crit}$).

As one decreases the energy, the zero velocity curves $r=r_p$ and $r = r_a$ come closer together.  Eventually, one will coincide with $r_{\rm crit}$.  
This occurs at the critical energy $E_{\rm crit}$ and brings about the bifurcation in the system.  For $E < E_{\rm crit}$,
the points $Y = \pm r_{\rm crit}$ both lie in the forbidden region, thus a transition from the left half to right half plane is no 
longer possible.   If the apoapsis zero-velocity curve $r=r_a$ reaches $r_{\rm crit}$ before the periapsis zero-velocity curve $r=r_p$ does, 
the particle follows 
periapsis librations (i.e. librations around periapsis on the right half plane) in the eccentric frame.  Alternatively,
if the periapsis zero-velocity curve $r_p$ reaches $r_{\rm crit}$ first, the particle follows apoapsis librations.

\subsection{The Route to Periapsis Librations}

\begin{figure}[ht!]
\begin{center}
\psfrag{T1}{(a) $E>>E_{\rm crit}$}
\psfrag{T2}{(b) $E\gtrapprox E_{\rm crit}$}
\psfrag{T3}{(c) $E = E_{\rm crit}$}
\psfrag{T4}{(d) $E \lessapprox E_{\rm crit}$}
\psfrag{T5}{(e) $E < E_{\rm crit}$}
\psfrag{T6}{(f) $E = E_{\rm circ}$}
\psfrag{B}{$\mathbf{\hat{B}}$}
\includegraphics[width=1.5in]{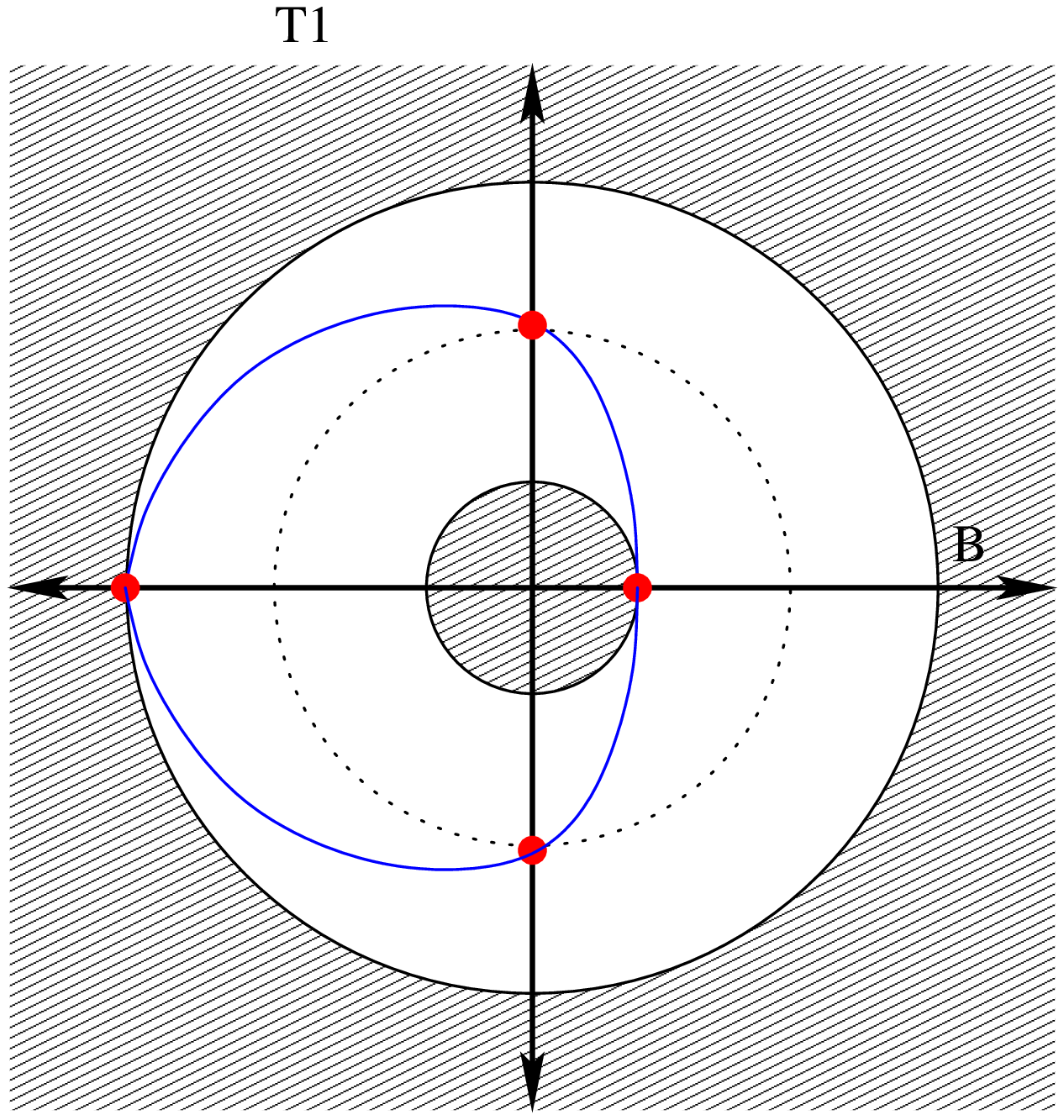}
\includegraphics[width=1.5in]{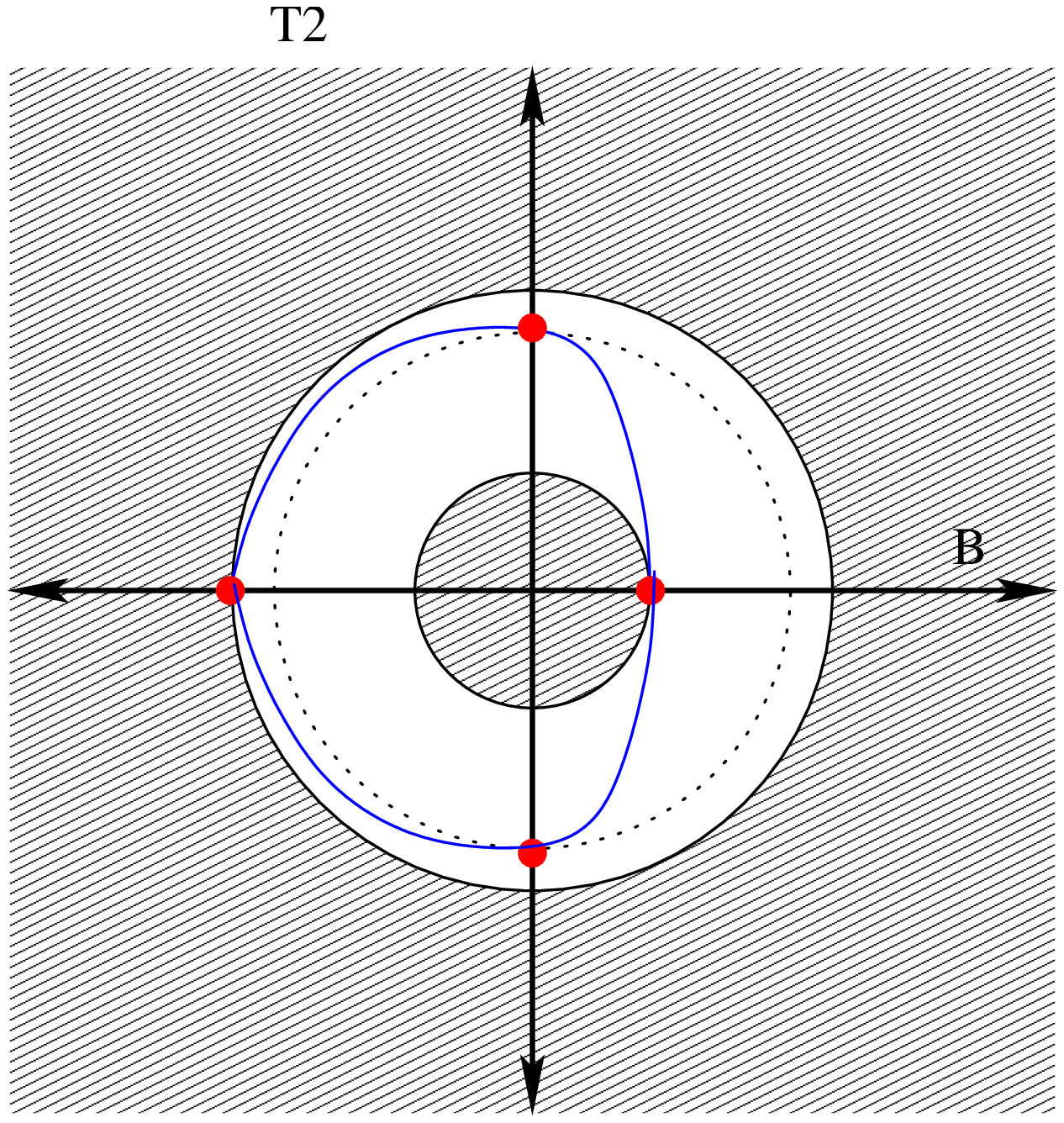}
\includegraphics[width=1.5in]{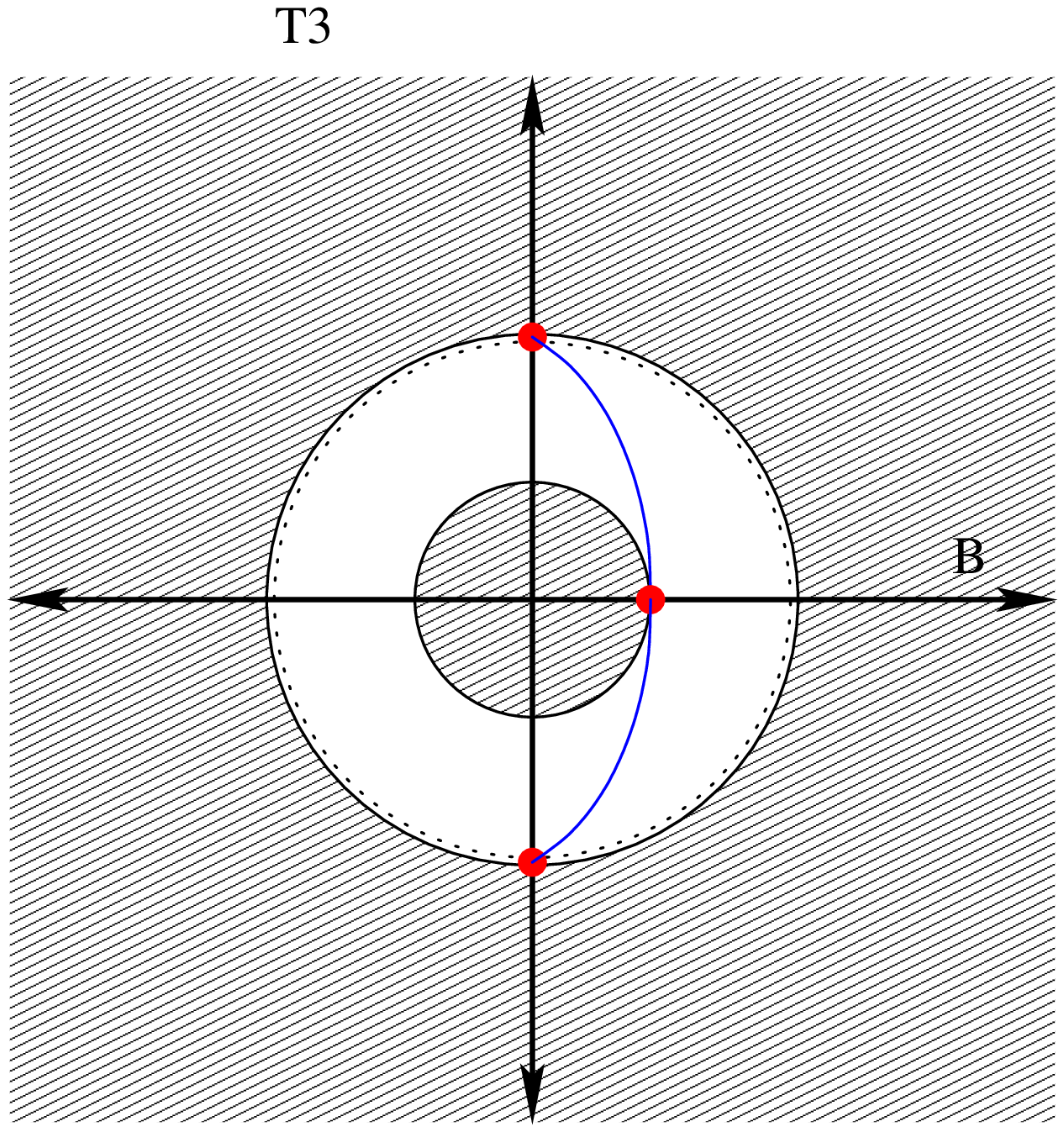}

\medskip

\includegraphics[width=1.5in]{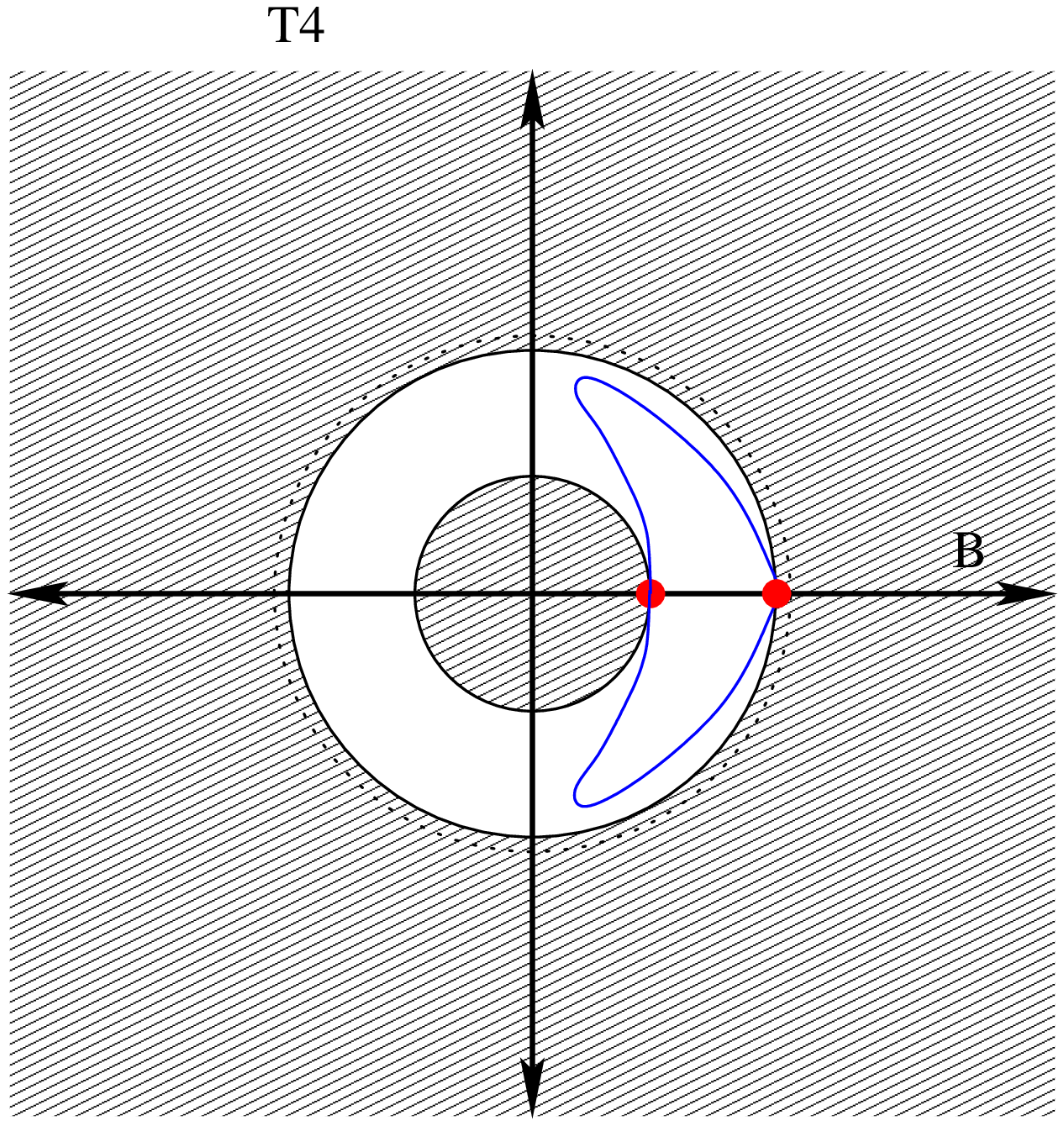}
\includegraphics[width=1.5in]{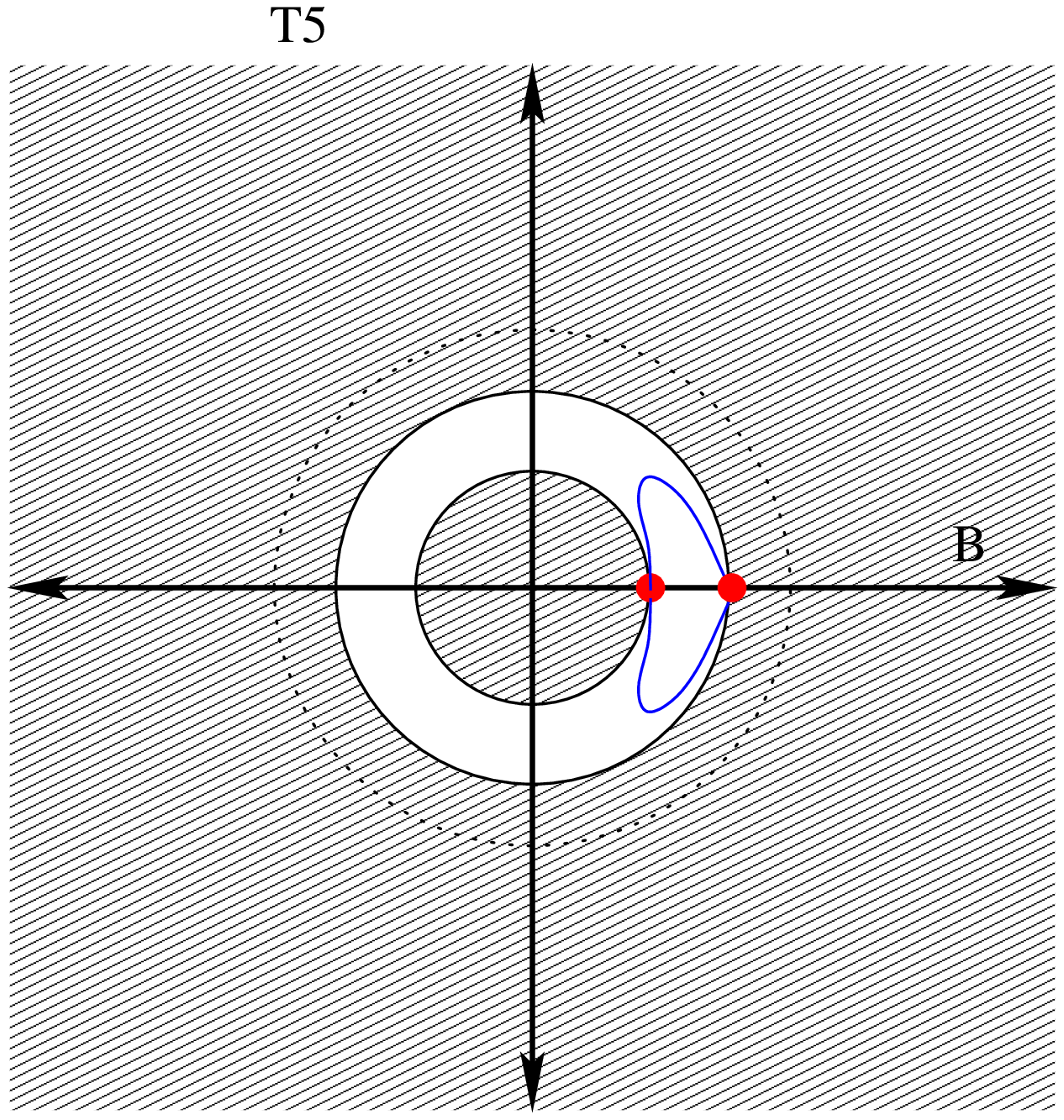}
\includegraphics[width=1.5in]{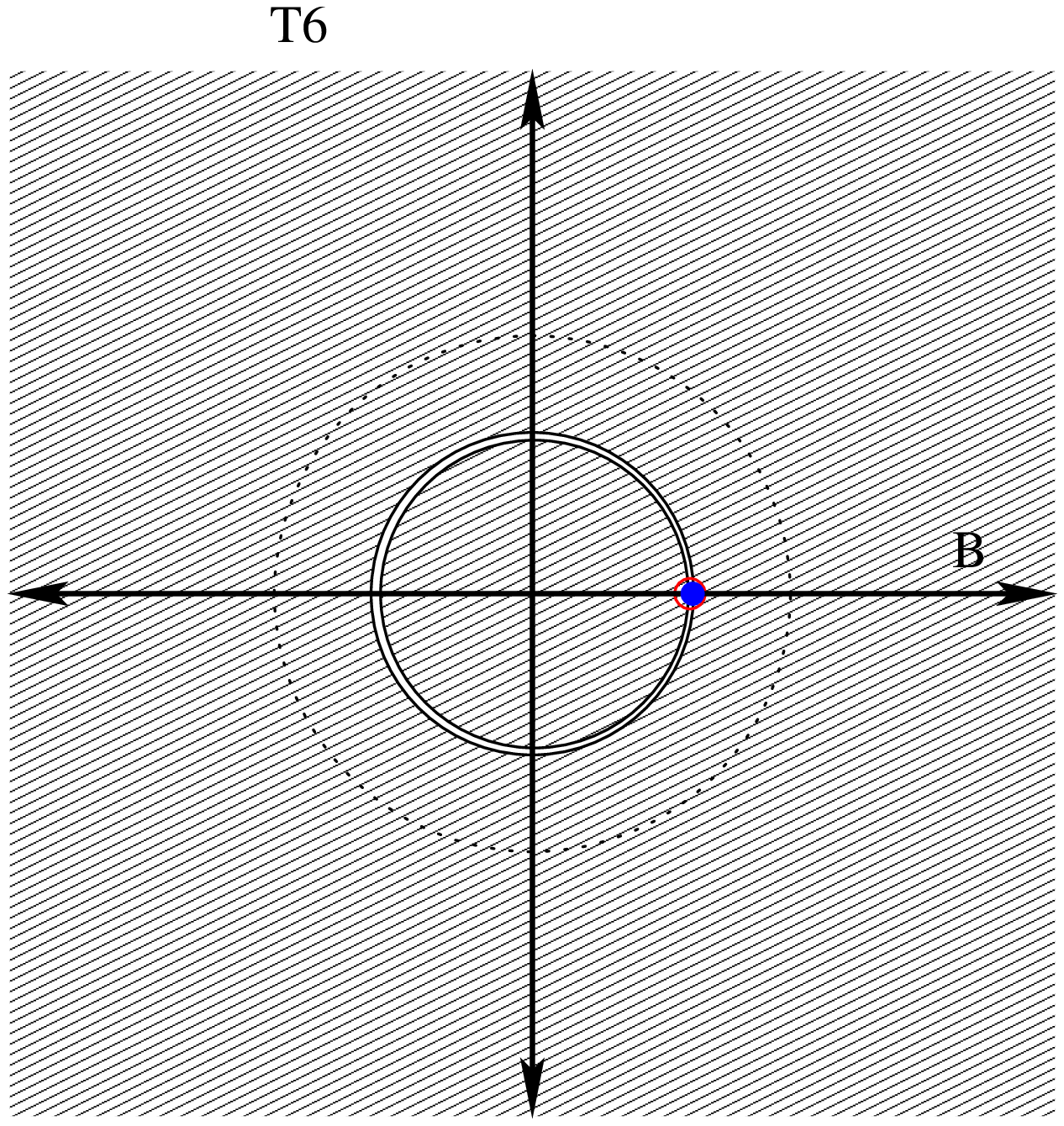}

\end{center}
\caption{The Route to Periapsis Librations}
\label{eccwaywardfig}
\end{figure}

We will first consider the case where it is the apoapsis radius that coincides with $r_{\rm crit}$ at the bifurcation energy
$E_{\rm crit}$.  This event brings about periapsis librations for all energies $E < E_{\rm crit}$.  
In Fig. \ref{eccwaywardfig} and Fig.  \ref{eccleewardfig}, the path of the particle, for various values of energy,
is plotted {\it with respect to the eccentric frame}, i.e. the $X$-axis is coincident with the eccentricity vector
$\hat{\mathbf{B}}$.  As viewed from this nonuniformly rotating frame, the trajectory of the particle
makes closed orbits.

If $E >> E_{\rm crit}$, \eqref{ecc08} and \eqref{ecc09} produce a well-defined closed orbit in the eccentric frame,
as seen in Figure \ref{eccwaywardfig}a.

As $E$ approaches $E_{\rm crit}$ from above, the apoapsis radius slowly approaches the critical radius, and an orbit
such as the one seen in Figure \ref{eccwaywardfig}b is present.  Notice the left half of this orbit is nearly circular.
This presents some numerical difficulty if one discretizes the radius $r$ and not
the true anomaly $f$.  However, this difficulty can be overcome by analytically approximating the left half of the orbit
with an ellipse $r_{\rm approx}(f), \ f \in [-\pi/2, -\pi] \cup [\pi/2, \pi]$ fitted to the data points 
$r_{\rm approx}(\pm \pi/2) = r_{\rm crit}$ and $r_{\rm approx}(\pi) = r_a$.  

At $E=E_{\rm crit}$, the apoapsis radius and the critical radius coincide, as shown in
Figure \ref{eccwaywardfig}c.  The particle thus reaches the $Y$-axis of the 
eccentric frame at the precise
moment it reaches the zero velocity curve.  Recall that \eqref{ecc09} implies that $r$ is increasing in the upper half plane
and decreasing in the lower half plane for the case $h > 0$.  This bifurcation point is rather interesting, as one only has
a half orbit in the eccentric frame.  The motion begins at periapsis, but when it reaches the $Y$-axis,
i.e. apoapsis, it ``hops'' $\pi$-radians to the corresponding point on the lower half plane and then returns to periapsis.  To compensate
there is a corresponding $\pi$-radian hop in the argument of periapsis, so that the true polar angle $\theta$ is a continuous function of time.
This is allowed as the $\frac{d\omega}{df}$ equation, \eqref{ecc11}, is actually undefined for $\dot r = 0$.  This is permissible 
because $B(r_{\rm crit}) = 0$ exactly if $E = E_{\rm crit}$, i.e. the eccentricity vector actually vanishes at these endpoints,
and then reappears pointing in the opposite direction.

For $E < E_{\rm crit}$, one sees that $r_a < r_{\rm crit}$.  Thus the portal $X = 0, \ Y = \pm r_{\rm crit}$ from the right- to the left-
half plane lies in the forbidden region.  Motion is therefore constrained to the right-half, where periapsis librations arise
in the eccentric frame, Figure \ref{eccwaywardfig}d-\ref{eccwaywardfig}e.

Finally, at $E = E_{\rm circ}$, the periapsis and apoapsis radii coincide and the trajectory in the eccentric frame 
degenerates to a single point $X = r_{\rm crit}, \ Y = 0$, as seen in
Figure \ref{eccwaywardfig}f.  The eccentric frame now rotates at a uniform
rate and a circular orbit is present in the actual inertial space.

\subsection{The Route to Apoapsis Librations}

\begin{figure}[ht!]
\begin{center}
\psfrag{T1}{(a) $E>>E_{\rm crit}$}
\psfrag{T2}{(b) $E\gtrapprox E_{\rm crit}$}
\psfrag{T3}{(c) $E = E_{\rm crit}$}
\psfrag{T4}{(d) $E \lessapprox E_{\rm crit}$}
\psfrag{T5}{(e) $E < E_{\rm crit}$}
\psfrag{T6}{(f) $E = E_{\rm circ}$}
\psfrag{B}{$\mathbf{\hat{B}}$}
\includegraphics[width=1.5in]{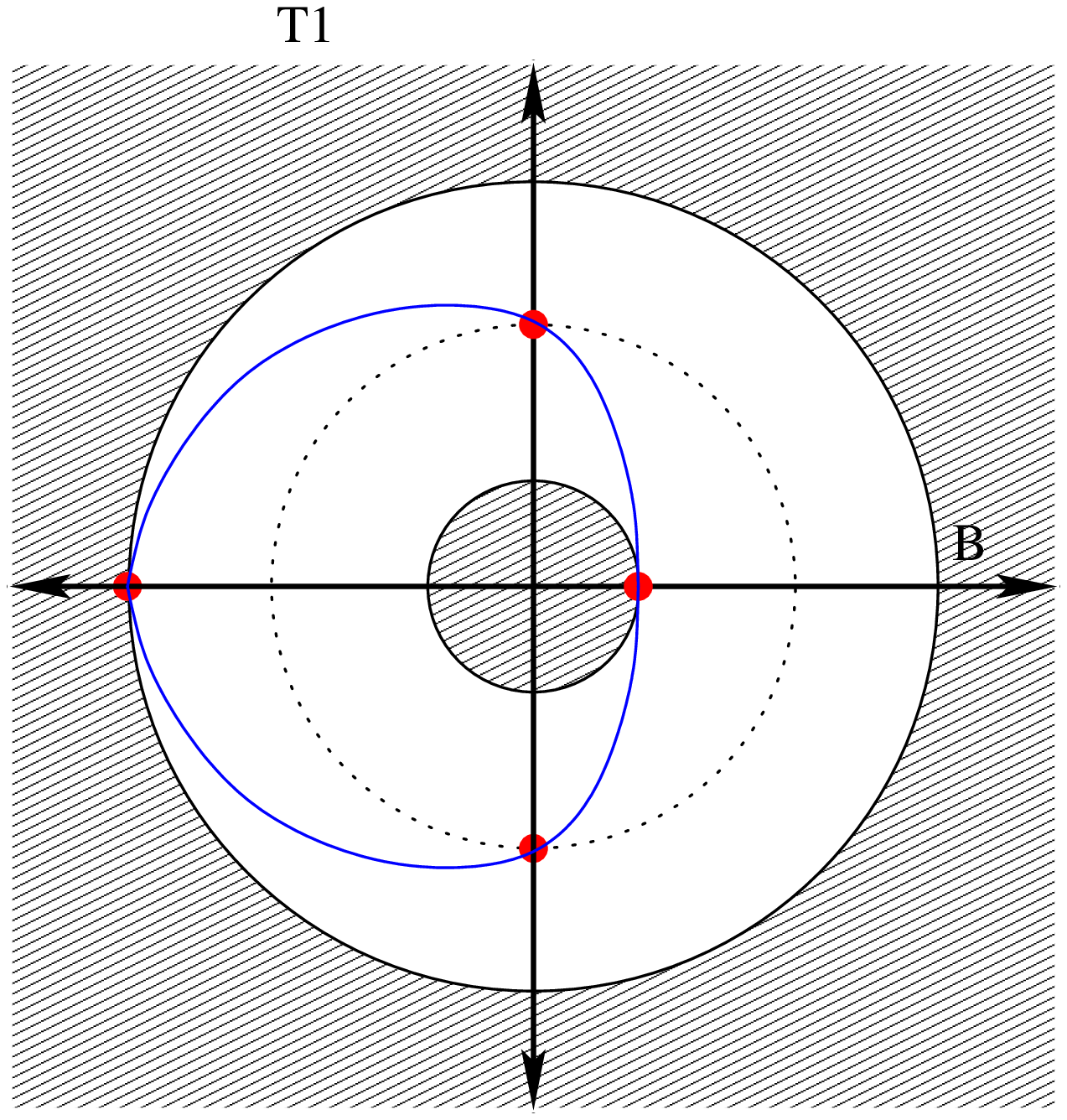}
\includegraphics[width=1.5in]{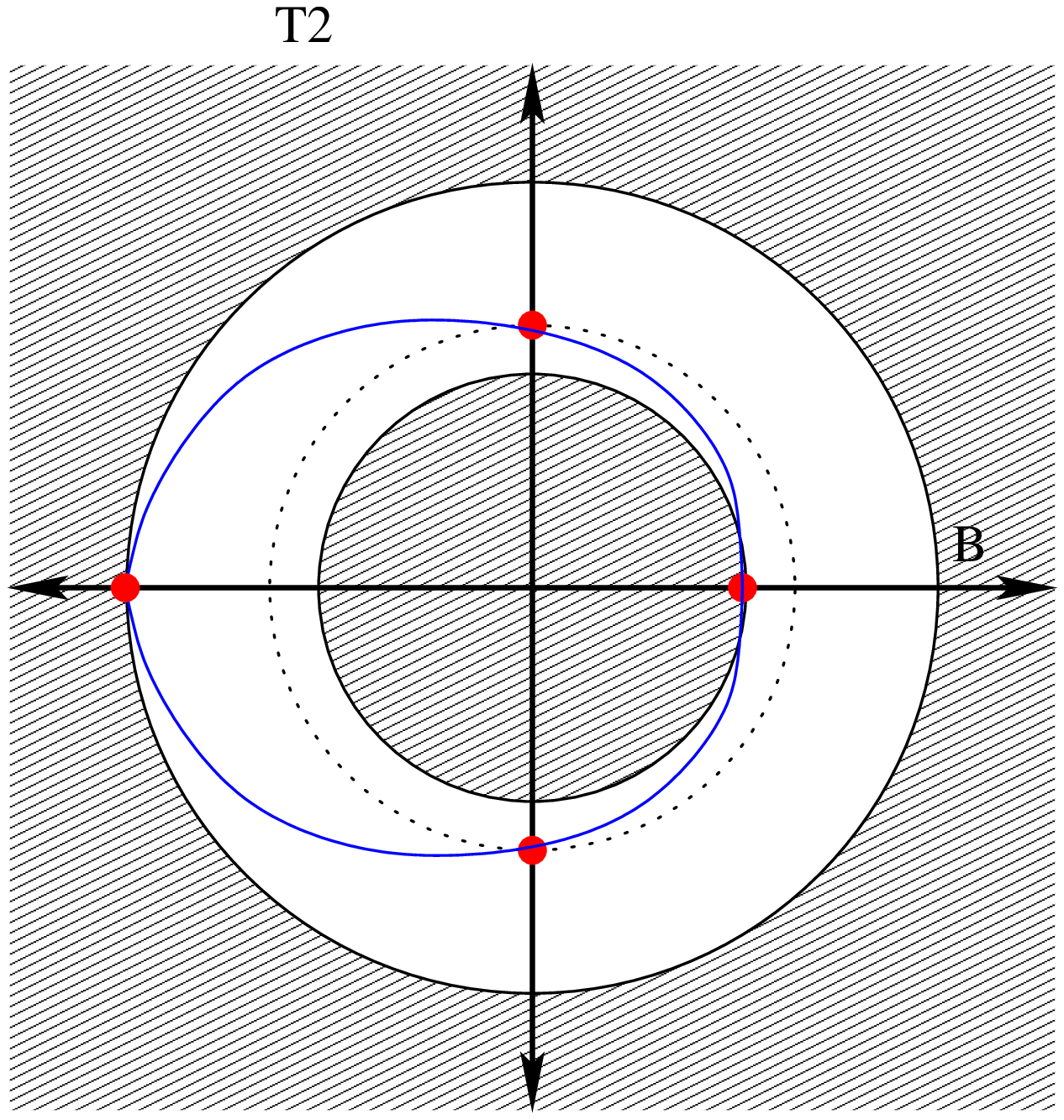}
\includegraphics[width=1.5in]{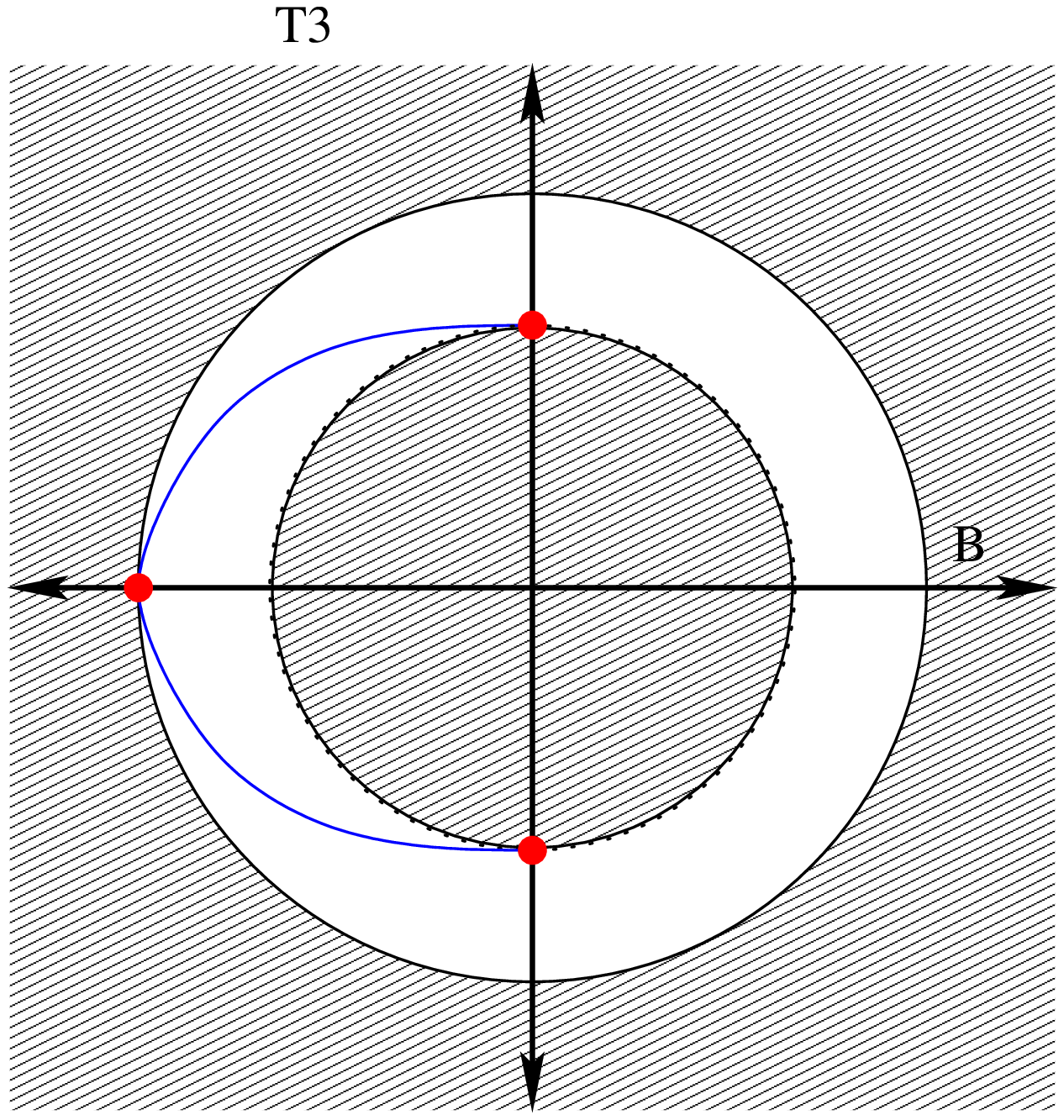}

\medskip

\includegraphics[width=1.5in]{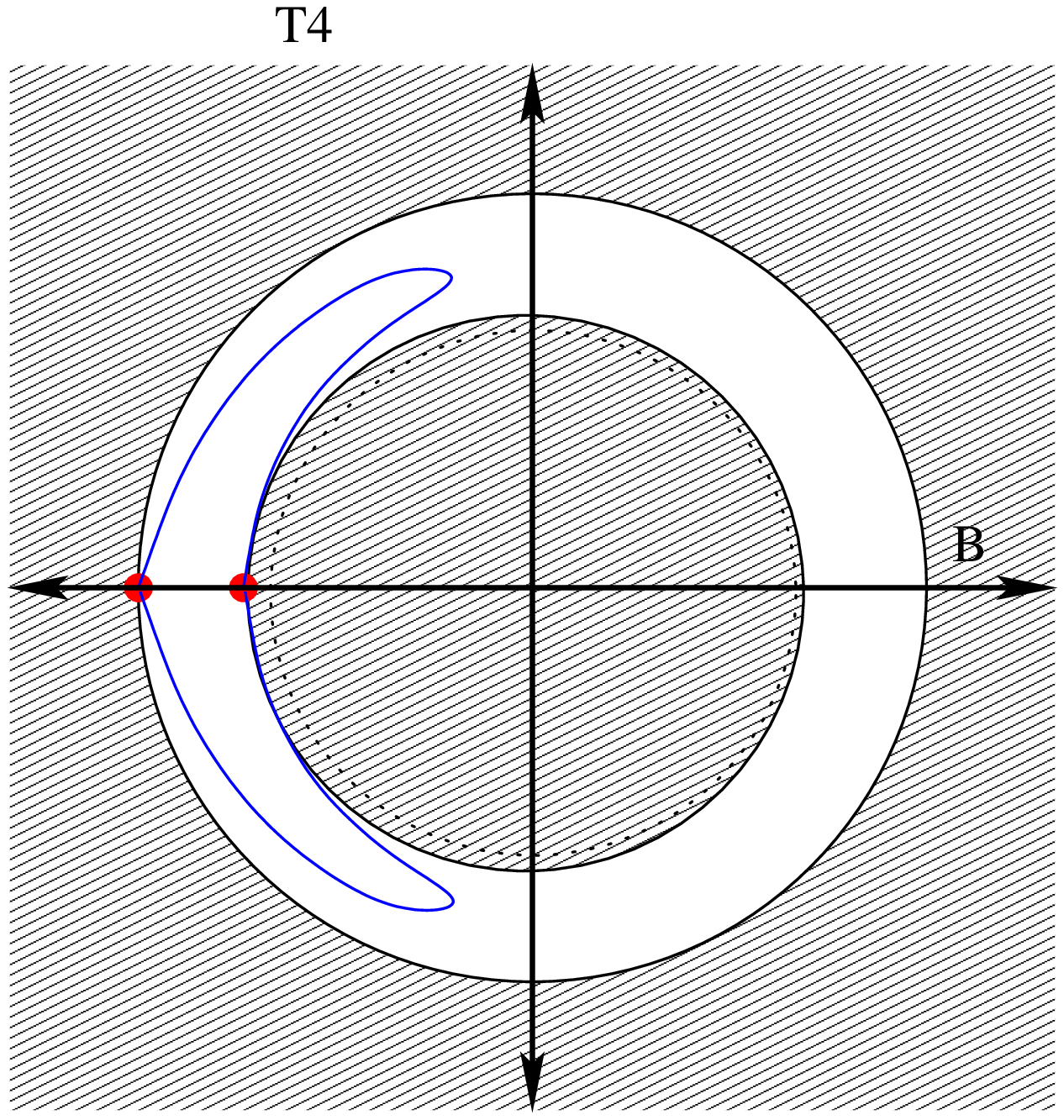}
\includegraphics[width=1.5in]{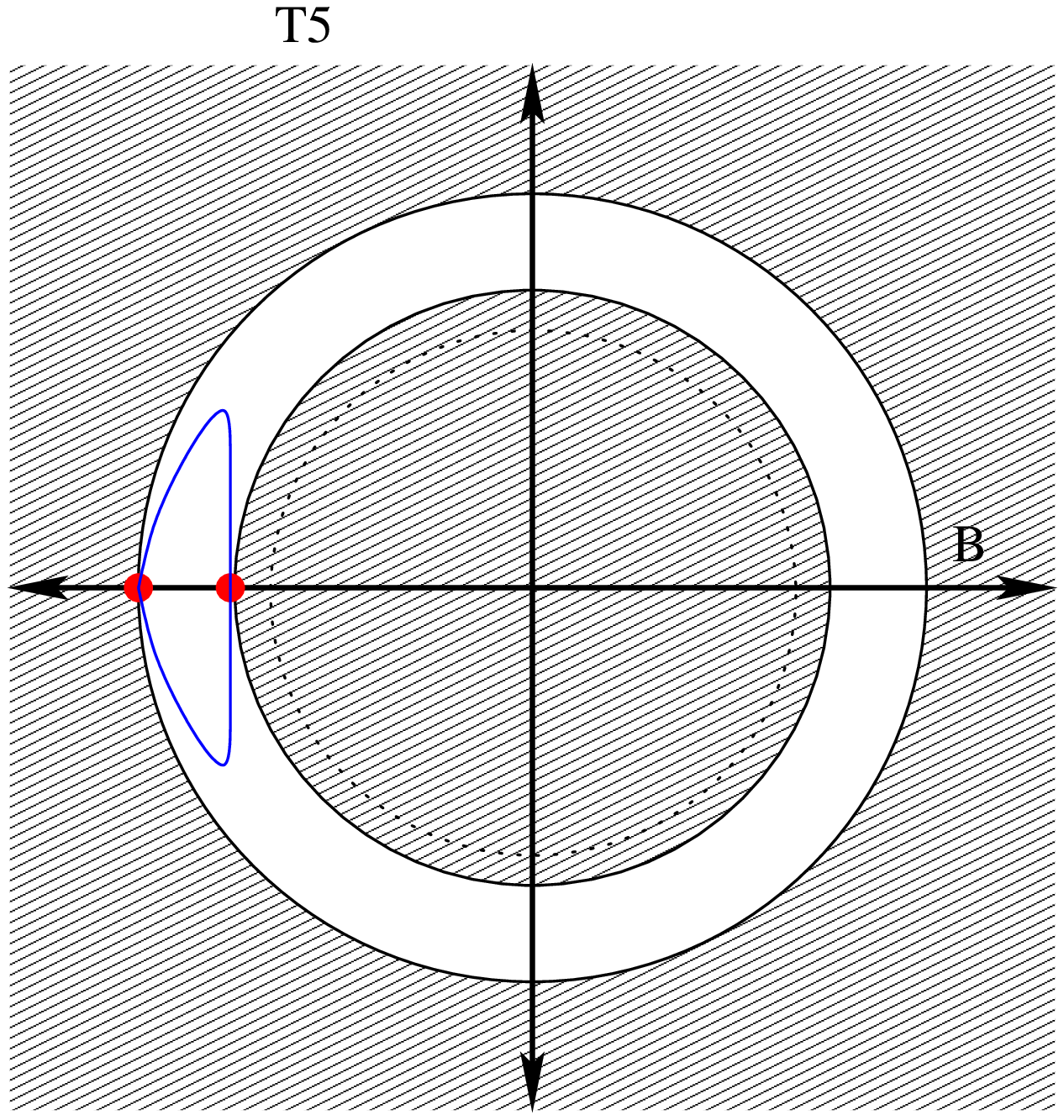}
\includegraphics[width=1.5in]{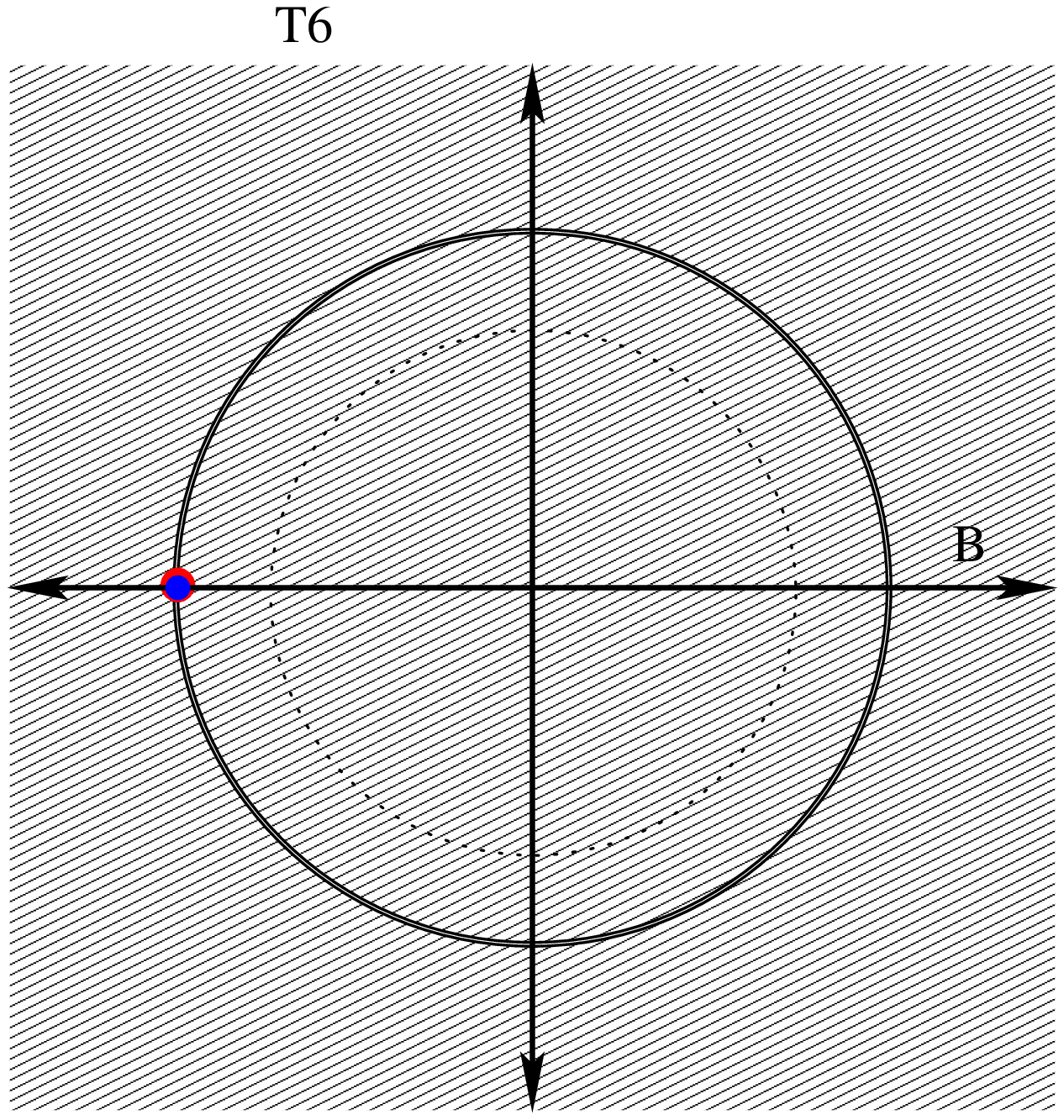}

\end{center}
\caption{The Route to Apoapsis Librations}
\label{eccleewardfig}
\end{figure}

As one sees from Figure \ref{eccleewardfig}, the case where the periapsis radius and the critical radius coincide at the bifurcation
energy $E = E_{\rm crit}$ leads to apoapsis librations in the left half plane.

An easy test to determine whether the librations will be periapsis or apoapsis librations is as follows:
\beas
r_{\rm circ} < r_{\rm crit} & \Longrightarrow & \mbox{periapsis librations} \\
r_{\rm circ} > r_{\rm crit} & \Longrightarrow & \mbox{apoapsis librations}.
\eeas

Again one sees that there is a $\pi$-radian hop in both true anomaly $f$ and argument of periapsis $\omega$
at the bifurcation energy $E = E_{\rm crit}$.  Just before the bifurcation, the right half of the orbit (the half closest to periapsis)
is nearly
circular.

\section{Symmetry of the Rotation}

One can exploit the form of the dynamical equation for $\omega$ \eqref{ecc11} to reduce the numerical integration
to one over only one half of an orbit.  By examining the differential equation \eqref{ecc11}, one sees that
$\omega'(f)$ depends only upon the radial coordinate $r$.  Due to the periodicity of the orbit, we have that 
$\omega'(f)$ is $2\pi$ periodic.  Moreover, for $f \in [\pi, 2\pi]$, we have that $\omega'(f) = \omega(2\pi - f)$, since the
orbits in the eccentric frame are symmetric with respect to the $x$-axis.

\subsection{Circulations}

Prior to the bifurcation ($E > E_{\rm crit}$) the trajectory makes closed circulations in the eccentric frame.  During 
the circulations, there is a secular growth in the argument of periapsis $\omega$.  Define the following:
\[
\tilde \Omega = \int_0^f \frac{d\omega}{df}(r(\tilde f)) \ d \tilde f \qquad f \in [0, \pi],
\]
such that $\tilde \Omega$ is the argument of periapsis restricted to the domain $f \in [0, \pi]$.  We will show that once one 
has $\tilde \Omega$, one can systematically find $\omega(f)$ for all future $f$, without integration.  

The condition that $\omega'(f) = \omega'(2\pi - f)$ for $f \in [\pi, 2\pi]$ suggests that the function $\omega(f)$ is odd 
with respect to the axes $f = \pi$ and $\omega = \omega(\pi)$ on the interval $[0, 2\pi]$.  Thus, given $\tilde \Omega$ (which 
we presume has been found by a numerical algorithm), one defines
\[
\Omega(f) = \left\{ \begin{array}{ll}
\tilde \Omega(f) & f \in [0, \pi] \\
2 \tilde \Omega(\pi) - \tilde \Omega(2\pi-f) \qquad & f \in (\pi, 2\pi]
\end{array} \right. .
\] 
The net secular growth in $\omega(f)$ over one nominal orbit $0 \le f \le 2\pi$ is given by
\[
\Delta \Omega = \Omega(2\pi).
\]
$\omega(f)$ can subsequently be found by applying the following:
\[
\omega(f) = n\Delta \Omega + \Omega( f \bmod 2 \pi),
\]
where $n$ is the orbit number, i.e. $n=0$ if $f \in [0, 2\pi]$, $n=1$ if $f \in [2\pi, 4\pi]$, etc.

\section{The Hernquist-Newton Potential}

To illustrate the theory in the context of a modern problem,
we will consider motion of a particle (star) in a spherical
galaxy, modelled with the Hernquist potential, with a central black
hole.  These results could be similarly applied to a 
black hole at the center of a globular cluster, or various
other astrophysical configurations that yield  spherical 
or azimuthal symmetry.  In this context, the central
black hole provides a classical point potential, but no general
relativistic effects are included.

\subsection{Galactic Halos with Central Black Holes}

The Hernquist potential has achieved some acclaim in recent years for its
ability to analytically model galactic dark matter halos, see Hernquist (1990).  We
will consider here a coupling between the spherical Hernquist profile and a Newtonian point mass,
assumed to model a black hole at the center of the galaxy.  Some numerical modelling
of triaxial galaxies with central black holes has already been carried out, as in Poon \& Merrit (2004).

Let $\mu_{\rm BH}$ and $\mu_{\rm halo}$ be the gravitational parameters of the central black hole
and the galactic dark matter halo, respectively; and let $b$ be a length scale of the galaxy (so that
$M(b) = M_{\rm tot}/4$, see Hernquist (1990)).  Then the Hernquist-Newton potential can be written:
\[
U(r) = \frac{\mu_{\rm halo}}{R + b} + \frac{\mu_{\rm BH}}{R}.
\]
By defining:
\[
\mu_0 = \mu_{\rm halo} + \mu_{\rm BH} \qquad \tilde \mu = \frac{\mu_{\rm halo}}{\mu_{\rm halo} + \mu_{\rm BH}},
\]
the Hernquist-Newton potential can be recast into the following equivalent form:
\[
U(r) = \frac{\mu_0}{R} \left( 1 - \frac{\tilde \mu}{1 + R/b } \right)
\]
with associated Hamiltonian:
\be
\label{ecc29}
\mathscr{E} = \frac 1 2 \left[ \left( \frac{dR}{dT} \right)^2 + \frac{H^2}{R^2} \right] - \frac{\mu_0}{R}
\left( 1 - \frac{\tilde \mu}{1 + R/b } \right).
\ee
where $H = R^2 \frac{d\theta}{dT}$ is the angular momentum.  As this is a central force field, the angular momentum
and energy will be conserved quantities.  
Observe that when $\tilde \mu = 0$, the potential energy reduces to that of a Newtonian
point mass.  When $\tilde \mu = 1$, the potential energy is equivalent to the Hernquist potential.  
For $0 < \tilde \mu << 1$, the model represents a Newtonian point mass with a
surrounding ``Hernquist cloud'' and for $0 << \tilde \mu < 1$, we have the Hernquist potential with a relatively weak
point mass at the origin, which could be used to model a spherical Hernquist galaxy with a central black hole.

\subsection{Nondimensionalization}

Carrying out the following change of variables:
\[
R = r b \qquad T = \sqrt{ \frac{b^3}{\mu_0} } t,
\]
and thus, consequently,
\[
H = \sqrt{b \mu_0} h \qquad \mathscr{E} = \frac{\mu_0}{b} E,
\]
we can recast the Hamiltonian \eqref{ecc29} into the following form:
\be
\label{ecc14}
E = \frac 1 2 \left( \dot r^2 + \frac{h^2}{r^2} \right) - \frac{\mu(r)}{r},
\ee
where we define
\be
\label{ecc15}
\mu(r) = \left( 1 - \frac{\tilde \mu}{1+r} \right).
\ee
We have thus recast the Hernquist-Newton potential to a one-parameter family of potentials, with 
$\tilde \mu = 1$ corresponding to a the Hernquist potential and $\tilde \mu = 0$ corresponding
to a pure Newtonian point mass.

We note that the Hernquist-Newton potential is similar to analogous work on the Manev
problem, which considers a potential of the form $U(r) = A/r + B/r^2$.
In fact, work has been carried out for the anisotropic Manev problem,
which replaces the radial coordinate $r$ with an ``elliptic radius''
$m = \sqrt{\mu x^2 + y^2}$
(e.g., Craig et al. 1999, Diacu \& Santoprete 2001).  In this type of
potential, one obtains a large class of chaotic orbits as well as
nonchaotic orbits.

\subsection{Zero Velocity Curves}

Using the relationship for circular orbits \eqref{ecc16} and the critical energy condition \eqref{ecc17}, sample values
of circular radius and energy, critical energy, and the critical periapsis and apoapsis are shown
in Table \ref{eccherntab}, where we have taken $h = 0.1$.  The critical radius $r_{\rm crit}$ coincides with 
the critical energy periapsis radius $r^{\rm peri}_{\rm crit}$ in each case, so that the bifurcation always
leads to apoapsis librations.

\begin{table}[ht!]
\begin{center}
\begin{tabular}{|c||c|c|c|c|c|c|}
\hline
$\tilde \mu$ & $r_{\rm circ}$ & $E_{\rm circ}$ & $E_{\rm crit}$ & $r^{\rm peri}_{\rm crit}$ & $r^{\rm apo}_{\rm crit}$ \\
\hline
1 & 0.2500 & -0.7200 & -0.4524 & 0.1051 & 1.1932 \\
\hline
0.99 & 0.2274 & -0.7539 & -0.5000 & 0.1000 & 1.000 \\
\hline
0.95 & 0.1508 & -0.9372 & -0.7440 & 0.0820 & 0.4460 \\
\hline
0.90 & 0.0938 & -1.3206 & -1.1960 & 0.0647 & 0.1645 \\
\hline
\end{tabular}
\end{center}
\caption{Various physical quantities for $h=1$}
\label{eccherntab}
\end{table}

\begin{figure}[ht!]
\begin{center}
\includegraphics[width=4in]{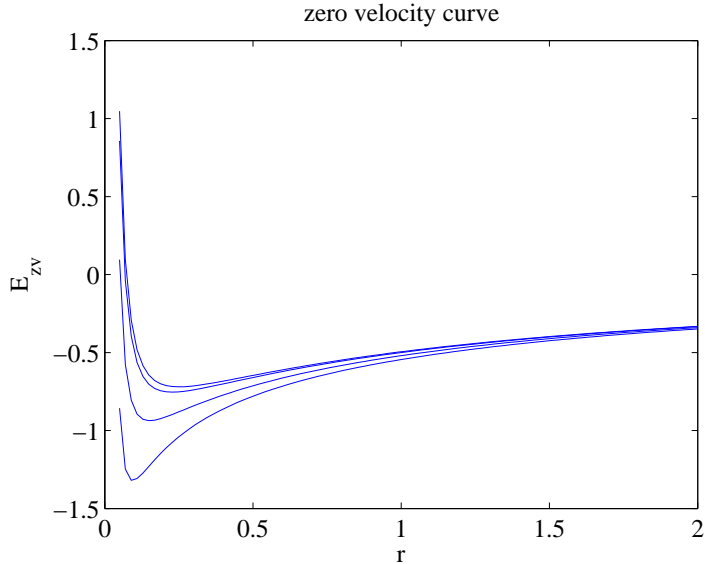}
\end{center}
\caption{Zero Velocity Curves with $h=1$ and, from top down, $\tilde \mu = 1, 0.99, 0.95, 0.9$}
\label{ecchern01zv}
\end{figure}

The zero-velocity curves are plotted in Fig. \ref{ecchern01zv} below for $h=0.1$ and for the
same values of $\tilde \mu$ as in Table \ref{eccherntab}.

\subsection{Orbits for $h=0.1$}

We examine a sample of orbits in the eccentric and inertial frames for various energies at 
$h=0.1$.  The zero velocity curve for this angular momentum is plotted in Fig. \ref{ecchern00} below.

\begin{figure}[ht!]
\begin{center}
\includegraphics[width=4in]{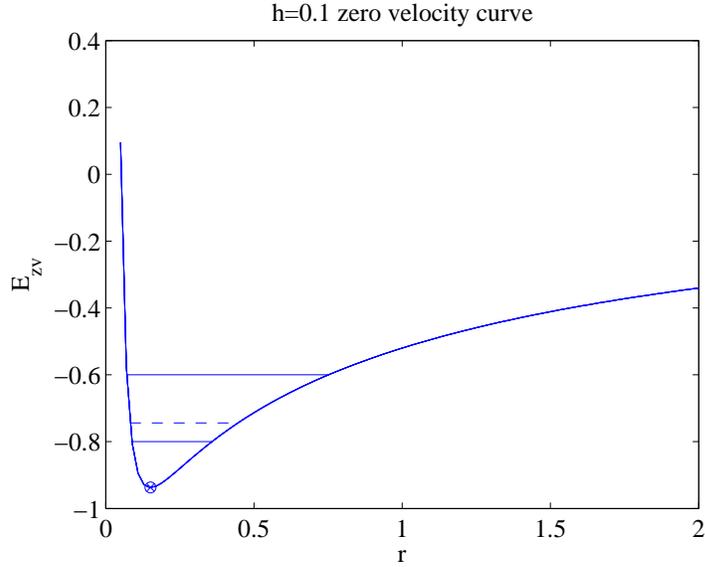}
\caption{Zero Velocity Curves and sample orbits at $h=0.1$; $E = -0.6, E_{\rm crit}, -0.8, E_{\rm circ}$}
\label{ecchern00}
\end{center}
\end{figure}

\begin{figure}[ht!]
\begin{center}
\includegraphics[width=2.6in]{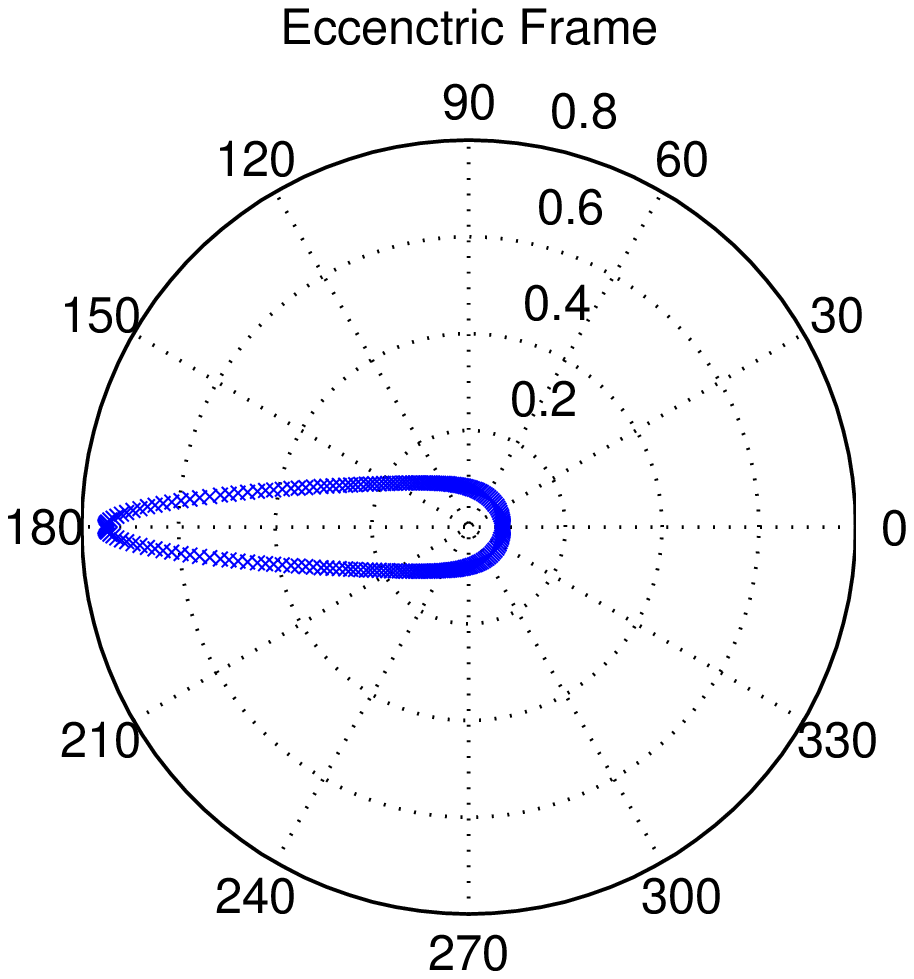}
\includegraphics[width=2.6in]{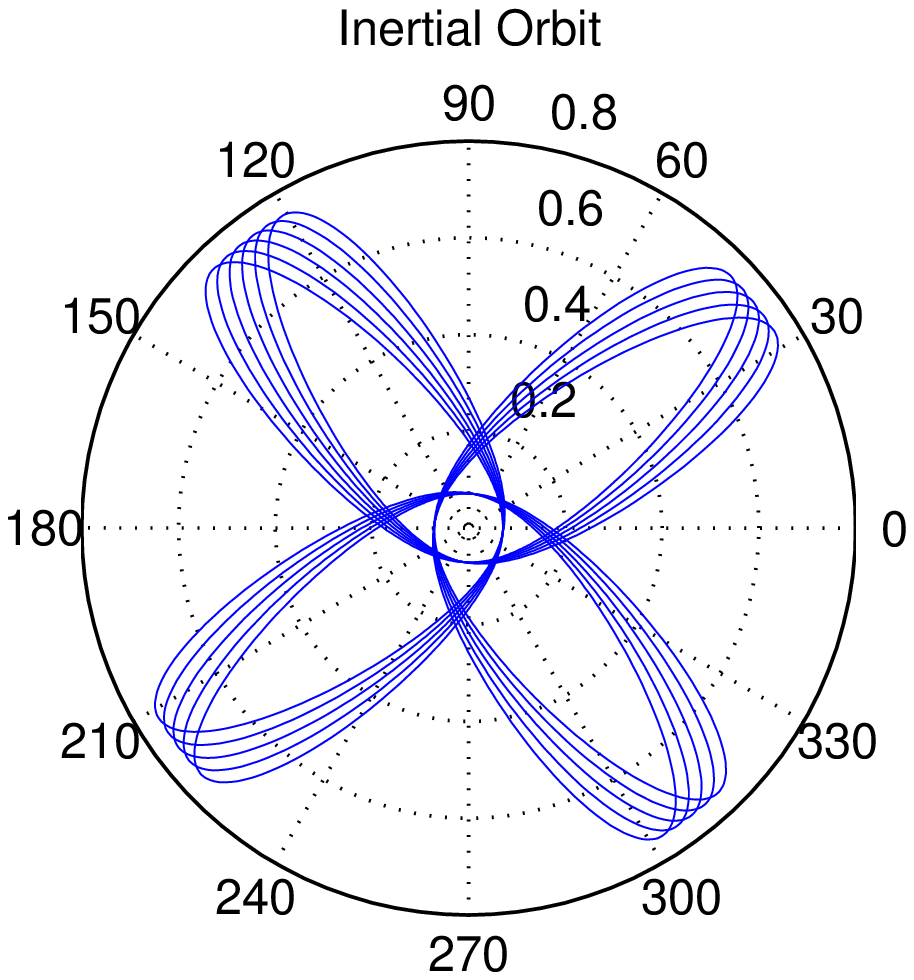}
\caption{$E = -0.6$ orbit in eccentric (left) and inertial (right) frames}
\label{ecchern12}
\end{center}
\end{figure}

For a sample orbit with $E > E_{\rm crit}$, we take $E=-0.6$.  The orbit as seen from the eccentric and inertial frames is shown
in Fig. \ref{ecchern12}.

\begin{figure}[ht!]
\begin{center}
\includegraphics[width=2.6in]{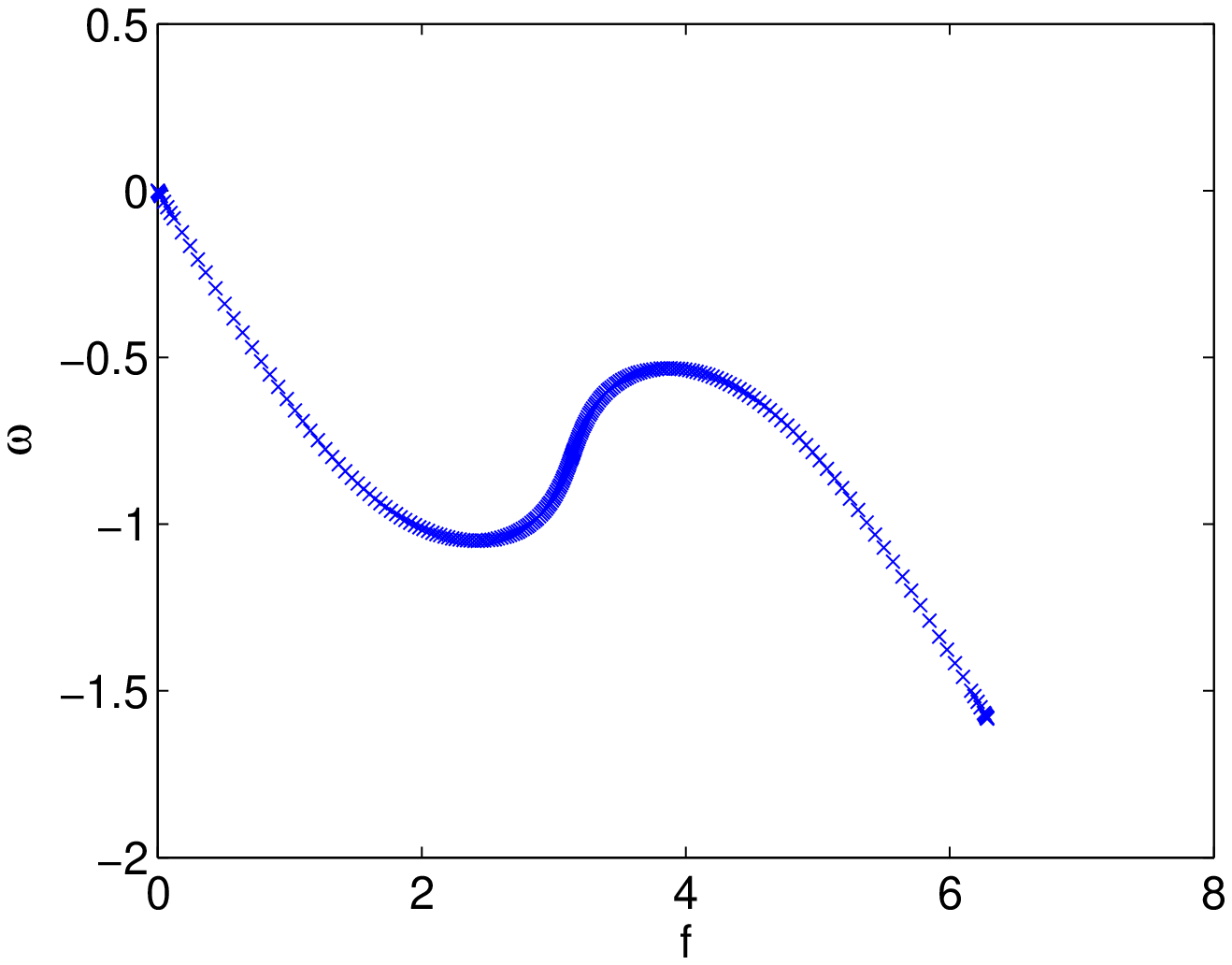}
\includegraphics[width=2.6in]{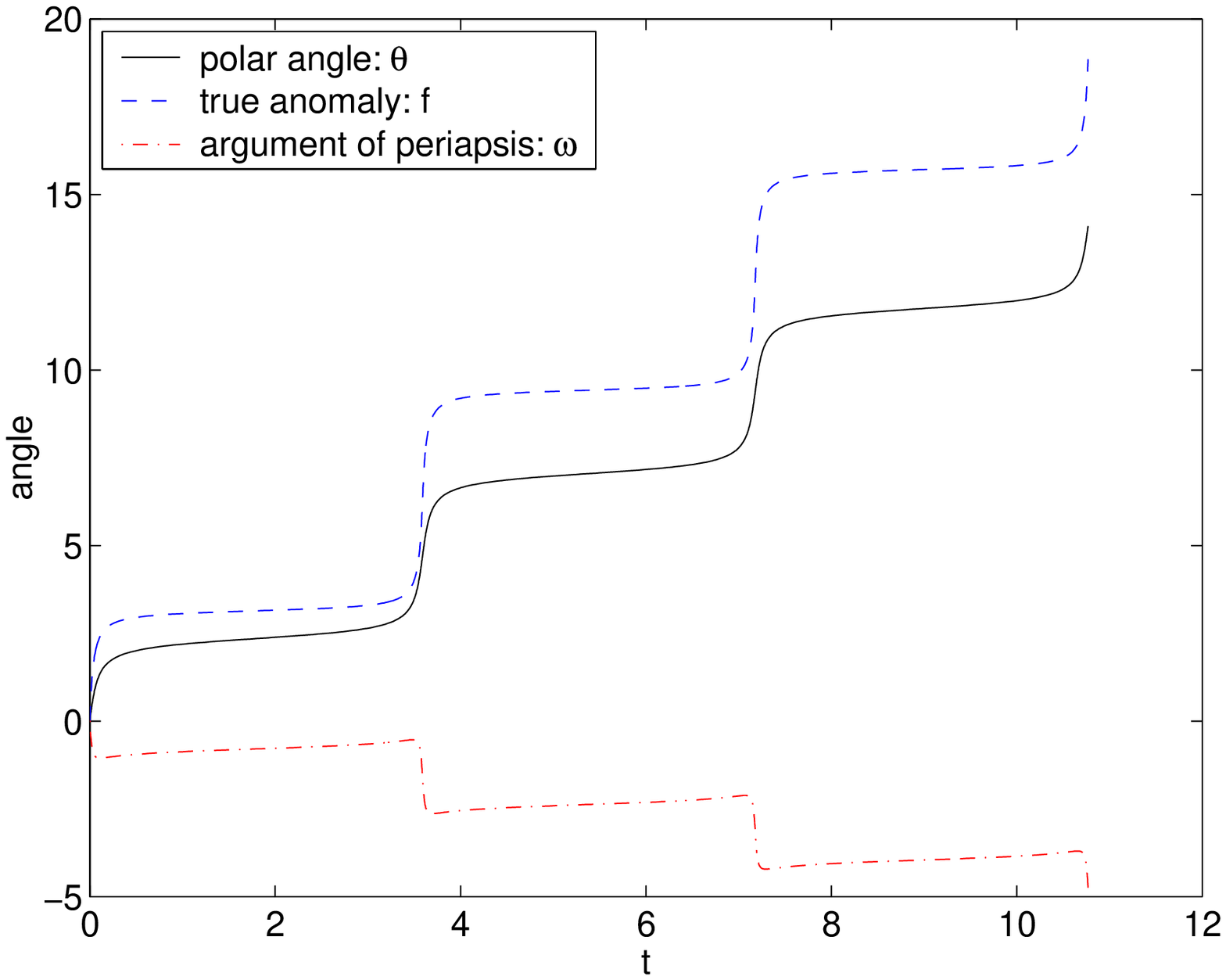}
\caption{$\omega(f)$ (left) and $\theta(t), \ \omega(t), \ f(t)$ (right) for $E = -0.6$}
\label{ecchern13}
\end{center}
\end{figure}

Upon integrating \eqref{ecc13}, one obtains $\omega(f)$, which can be seen for this orbit plotted in Fig. \ref{ecchern13}.  One sees
$\Delta \omega = \omega(2\pi) - \omega(0)$ is the \textit{turning angle} of the rosette.  For energies prior to (above) the
critical bifurcation energy, we find a secular retrograde rotation of the eccentric frame.  
$\theta(t)$, $\omega(t)$, and $f(t)$ are plotted against time over three standard orbits on the right.  One sees a secular
prograde growth in the inertial polar angle $\theta$, with turning angle $\Delta \theta = 2 \pi + \Delta \omega$.  
(Recall that $\Delta \omega < 0$).

\begin{figure}[ht!]
\begin{center}
\includegraphics[width=2.6in]{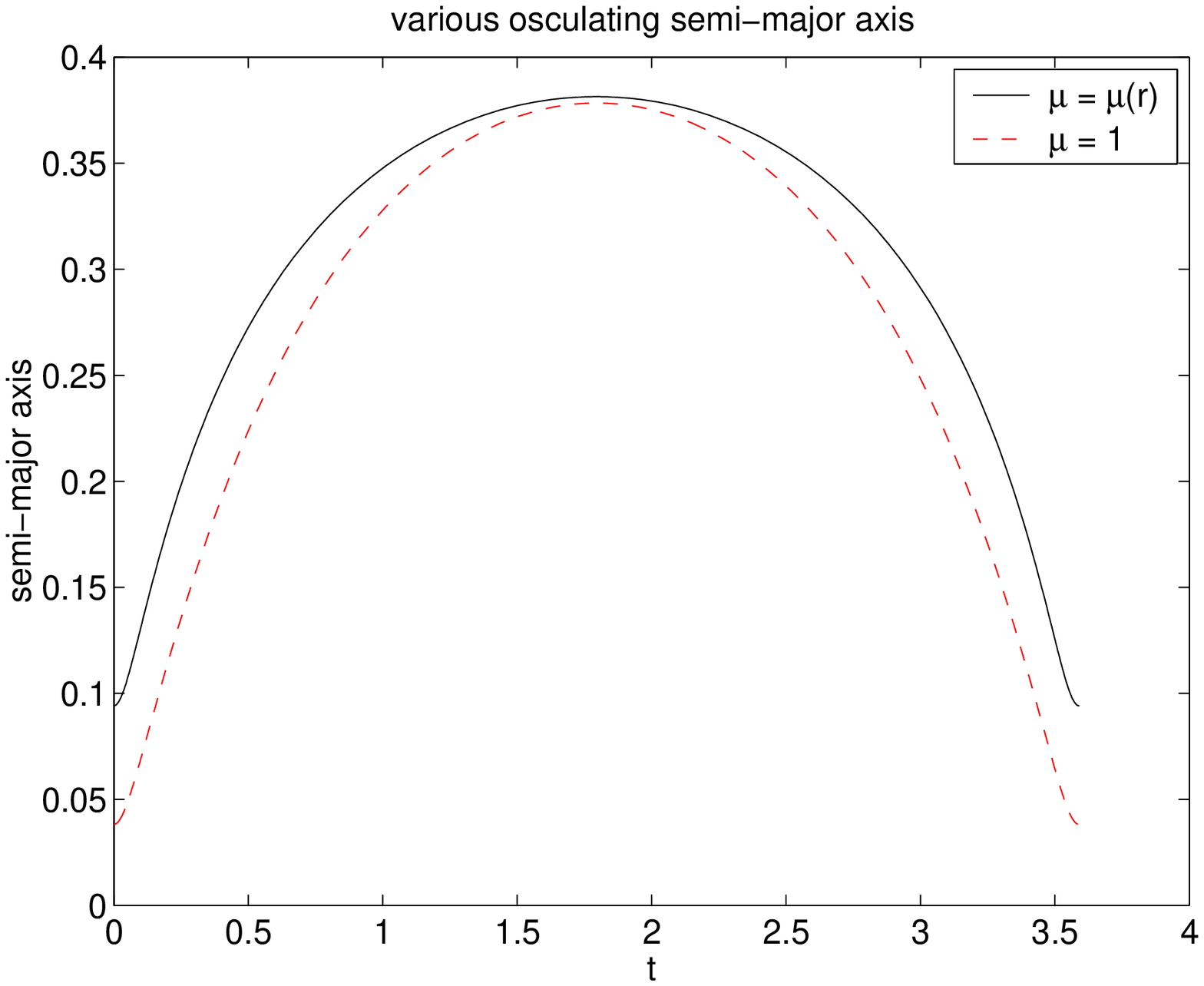}
\includegraphics[width=2.6in]{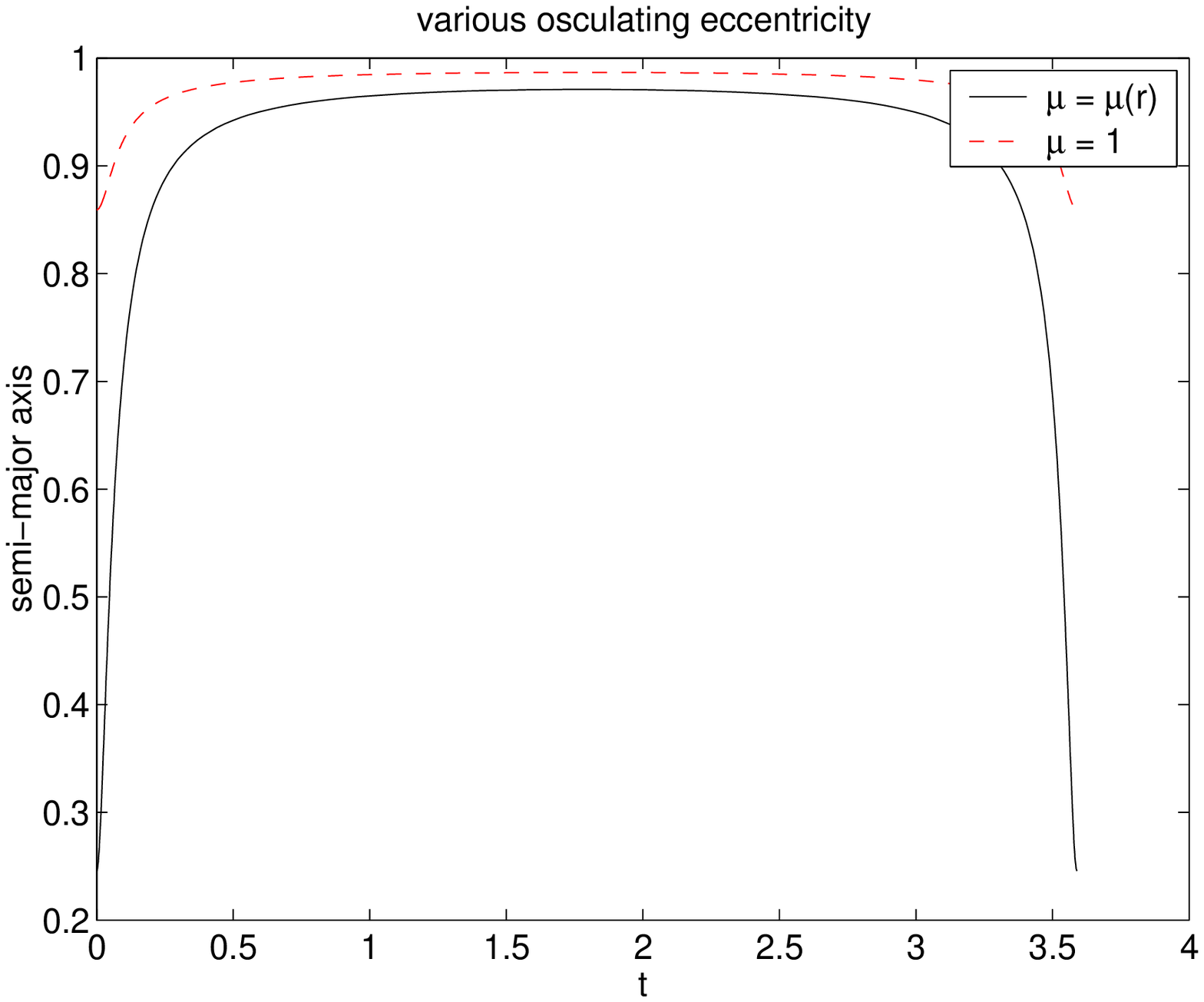}
\caption{Osculating orbital elements for $E = -0.6$}
\label{ecchern15}
\end{center}
\end{figure}

Finally, we compute the osculating semi-major axis and eccentricity vs. time (Fig. \ref{ecchern15}) 
over one nominal orbit (as seen from the eccentric frame).  The solid lines represent the osculating
elements as provided by the eccentric frame method, see \eqref{ecc25}-\eqref{ecc26}.  The dashed curves
are a standard set using the osculating orbital element transformation as defined by classical perturbation theory,
using $\mu_0 = 1$ for the ``planet'' mass, i.e. $\mu_0$ is the gravitational parameter of the \textit{total} 
halo mass \textit{plus} the mass of the central black hole.

\begin{figure}[ht!]
\begin{center}
\includegraphics[width=2.6in]{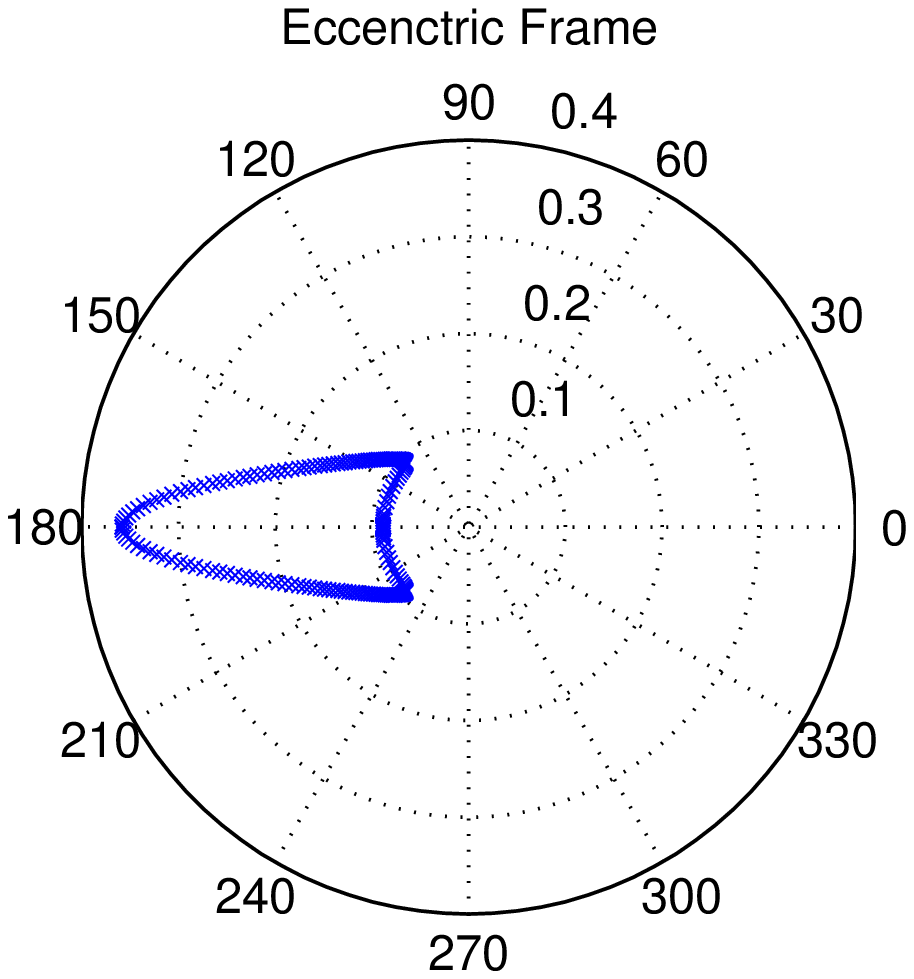}
\includegraphics[width=2.6in]{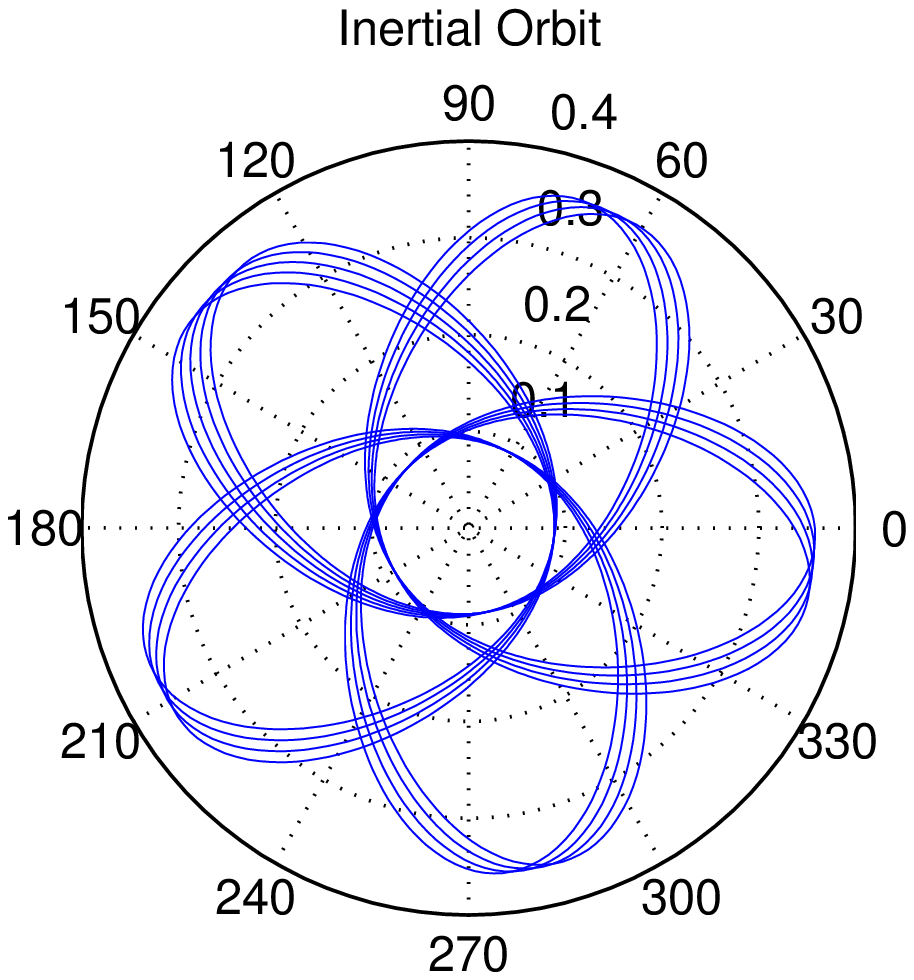}
\caption{$E = -0.8$ orbit in eccentric (left) and inertial (right) frames}
\label{ecchern22}
\end{center}
\end{figure}

For a sample orbit with $E < E_{\rm crit}$, we take $E=-0.8$.  The orbit as seen from the eccentric and inertial frames is shown
in Fig. \ref{ecchern22}.  Notice that the particle now makes librations in the eccentric frame.  The libration paths become smaller
and smaller until they degenerate to a single point at $E_{\rm circ}$, to be considered next.

\begin{figure}[ht!]
\begin{center}
\includegraphics[width=2.6in]{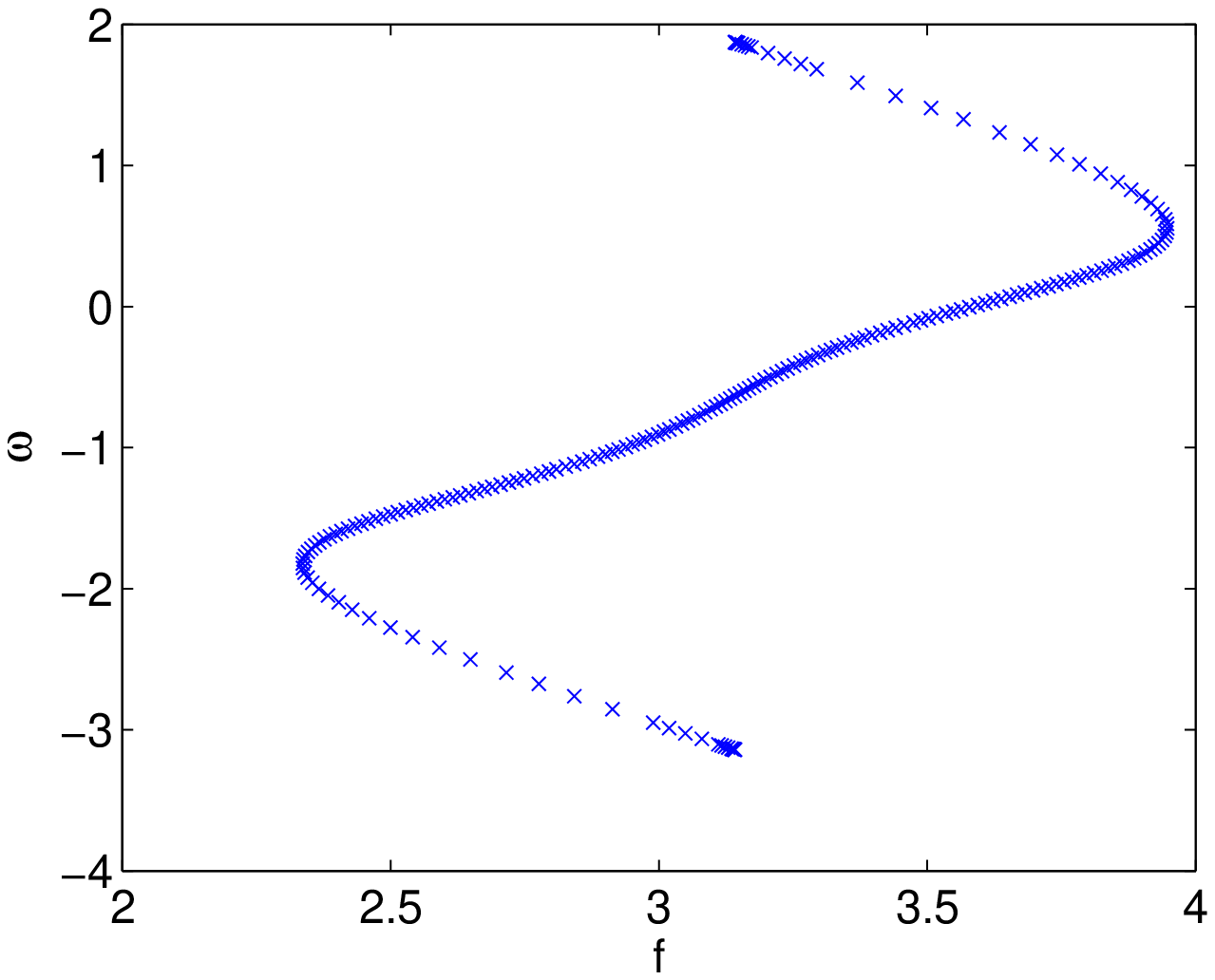}
\includegraphics[width=2.6in]{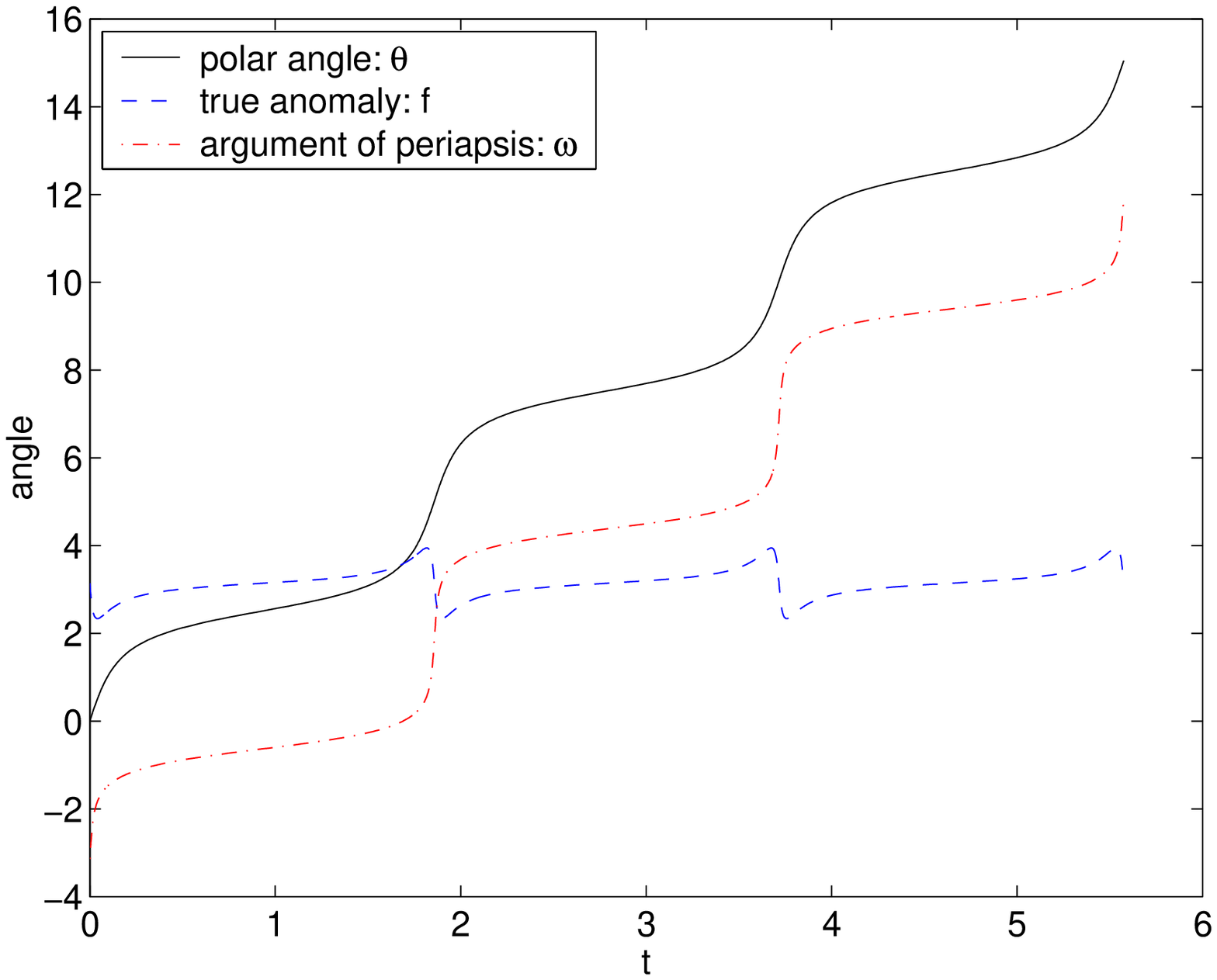}
\caption{$\omega(f)$ (left) and $\theta(t), \ \omega(t), \ f(t)$ (right) for $E = -0.8$}
\label{ecchern23}
\end{center}
\end{figure}

$\omega(f)$ for this orbit is plotted in Fig. \ref{ecchern23}.  Notice that $f$ librates around $f = \pi$
and there is secular prograde growth in $\omega$.  The turning angle is still given by
$\Delta \omega$.  For energies after (below) the
critical bifurcation energy, we find a secular prograde rotation of the eccentric frame.  
$\theta(t)$, $\omega(t)$, and $f(t)$ are plotted against time over three standard orbits on the right.  One now sees a secular
posigrade growth in the argument of periapsis, coupled with librations in true anomaly $f$.

\begin{figure}[ht!]
\begin{center}
\includegraphics[width=2.6in]{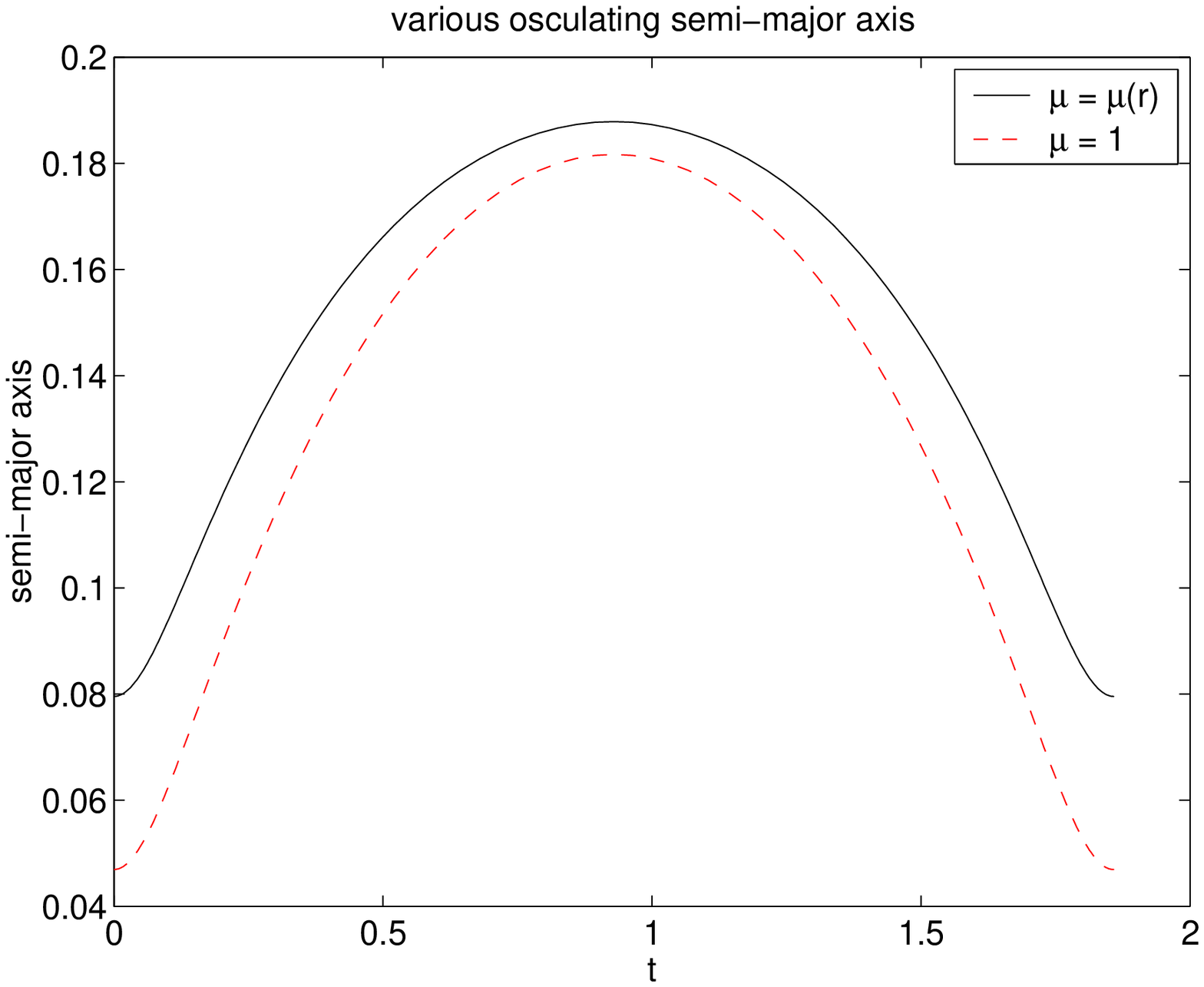}
\includegraphics[width=2.6in]{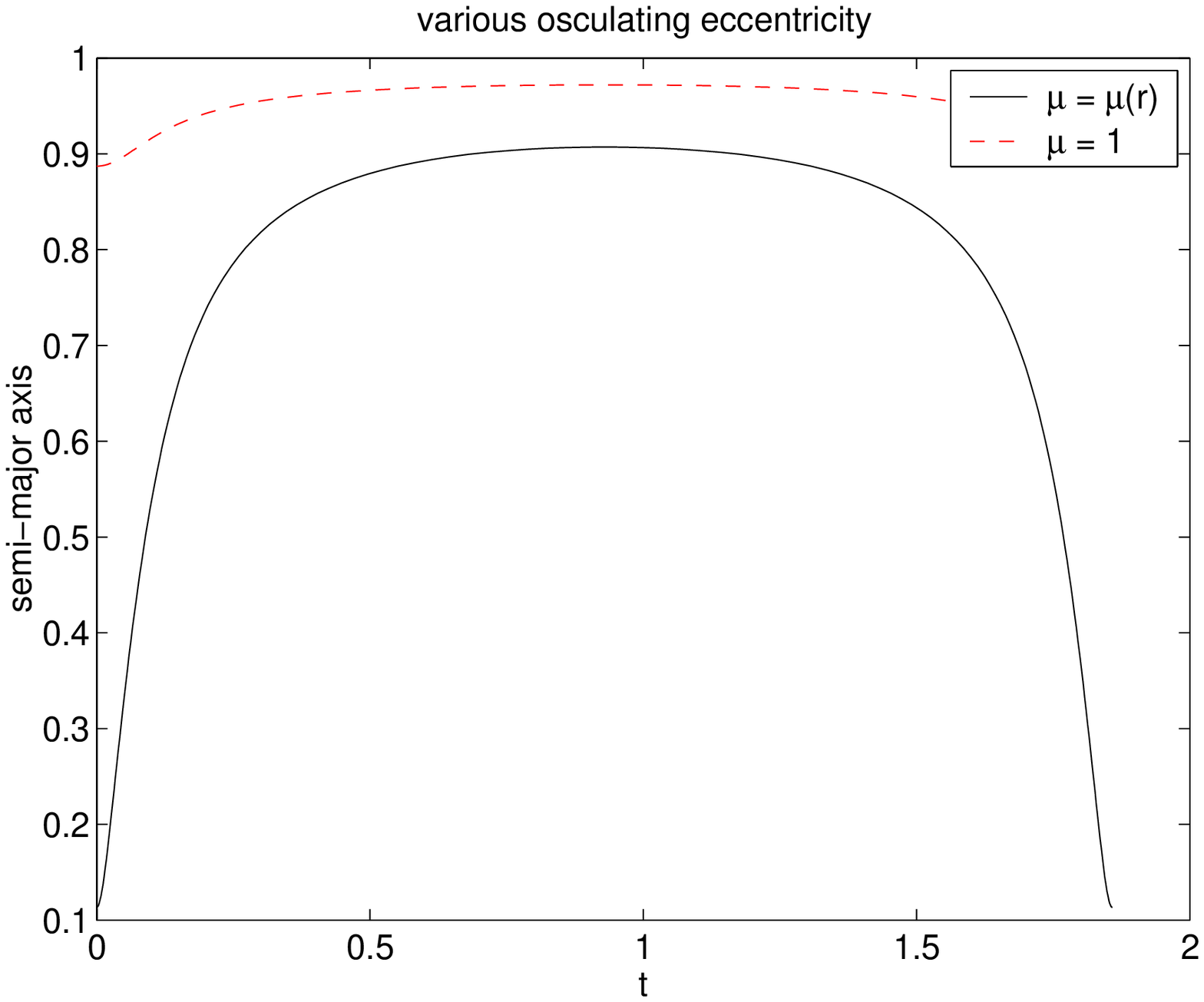}
\caption{Osculating orbital elements for $E = -0.8$}
\label{ecchern25}
\end{center}
\end{figure}

Finally, we compute the osculating semi-major axis  and eccentricity vs. time (Fig. \ref{ecchern25})
over one nominal orbit (as seen from the eccentric frame).  The solid lines represent the osculating
elements as provided by the eccentric frame method, see \eqref{ecc25}-\eqref{ecc26}.  As before, the dashed curves
are a standard set using the osculating orbital element transformation as defined by classical perturbation theory,
using $\mu_0 = 1$ for the ``planet'' mass, i.e. $\mu_0$ is the gravitational parameter of the \textit{total} 
halo mass \textit{plus} the mass of the central black hole.

\begin{figure}[ht!]
\begin{center}
\includegraphics[width=2.6in]{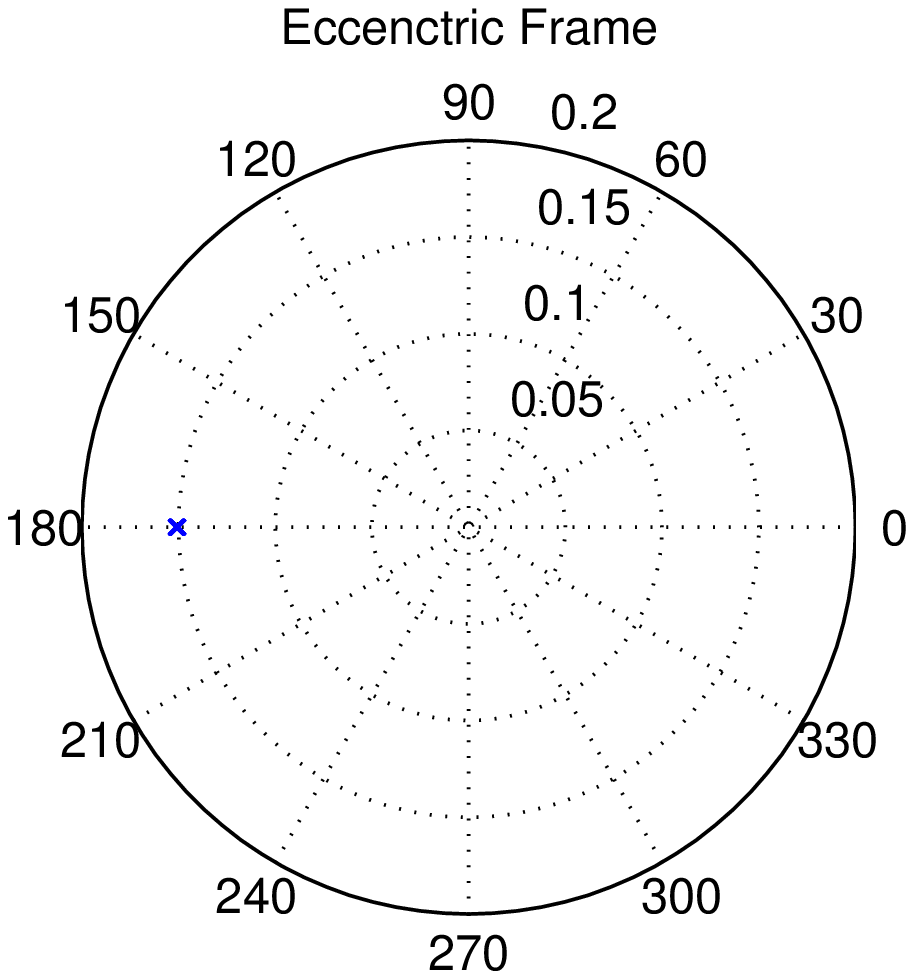}
\includegraphics[width=2.6in]{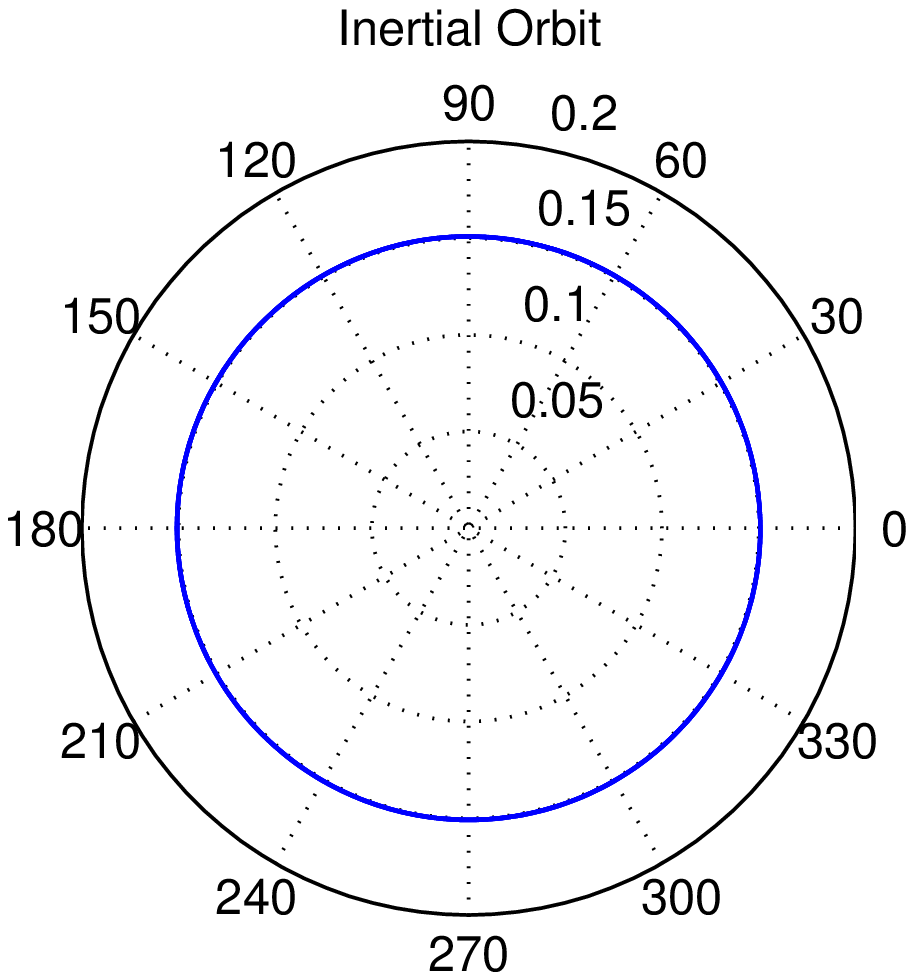}
\caption{$E = -0.9372$ orbit in eccentric (left) and inertial (right) frames}
\label{ecchern32}
\end{center}
\end{figure}

At the circular energy $E = E_{\rm circ}$, the orbit in the eccentric frame degenerates to a single point.
The orbit is circular in the inertial plane, Fig. \ref{ecchern32}.  The eccentric frame now precesses at a uniform rate.

$\omega(f)$ (left) and $\theta(t), \omega(t), f(t)$ are plotted below in Fig. \ref{ecchern33}.  Now $f$ is virtually constant
and there is uniform growth in $\omega$.

\begin{figure}[ht!]
\begin{center}
\includegraphics[width=2.6in]{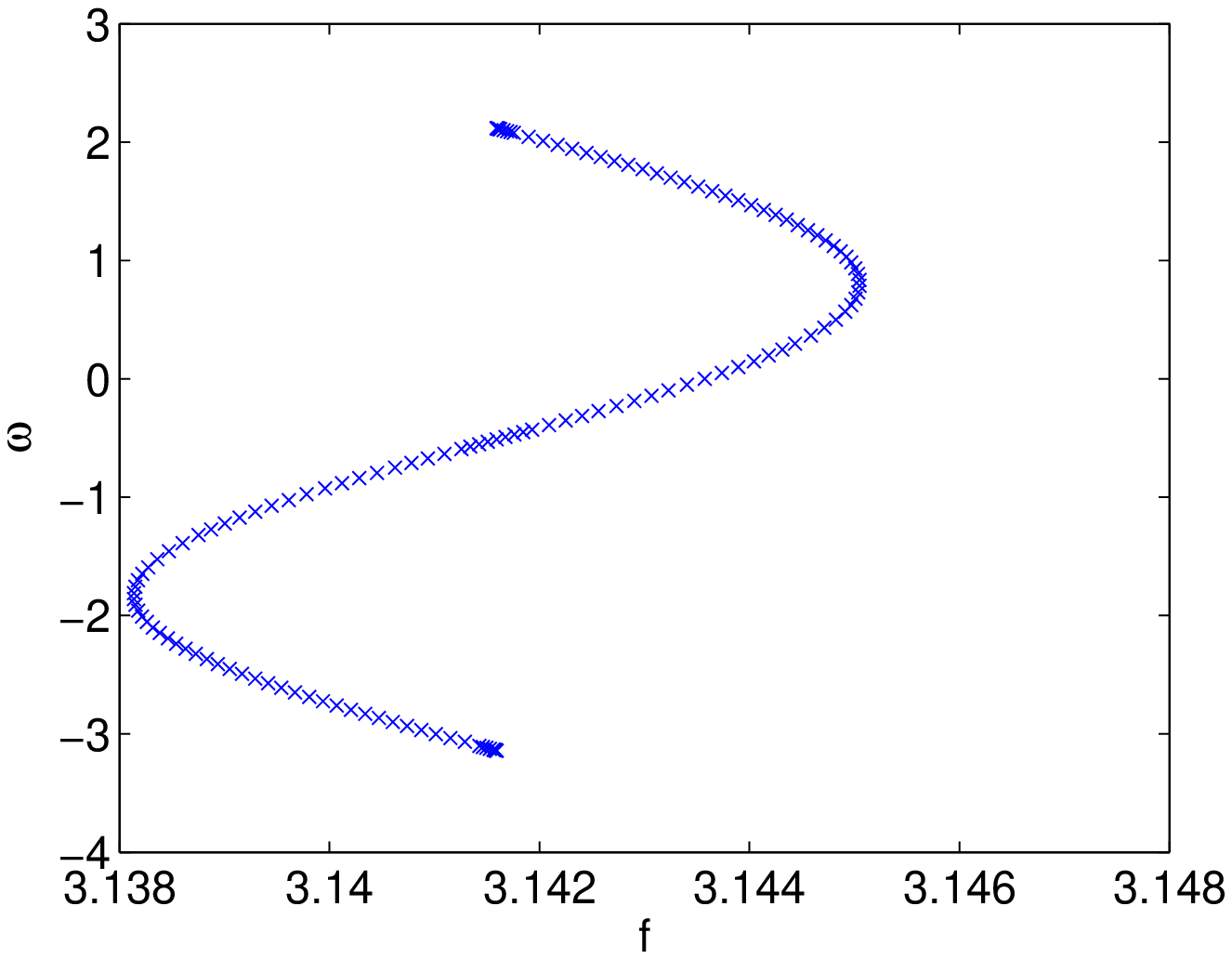}
\includegraphics[width=2.6in]{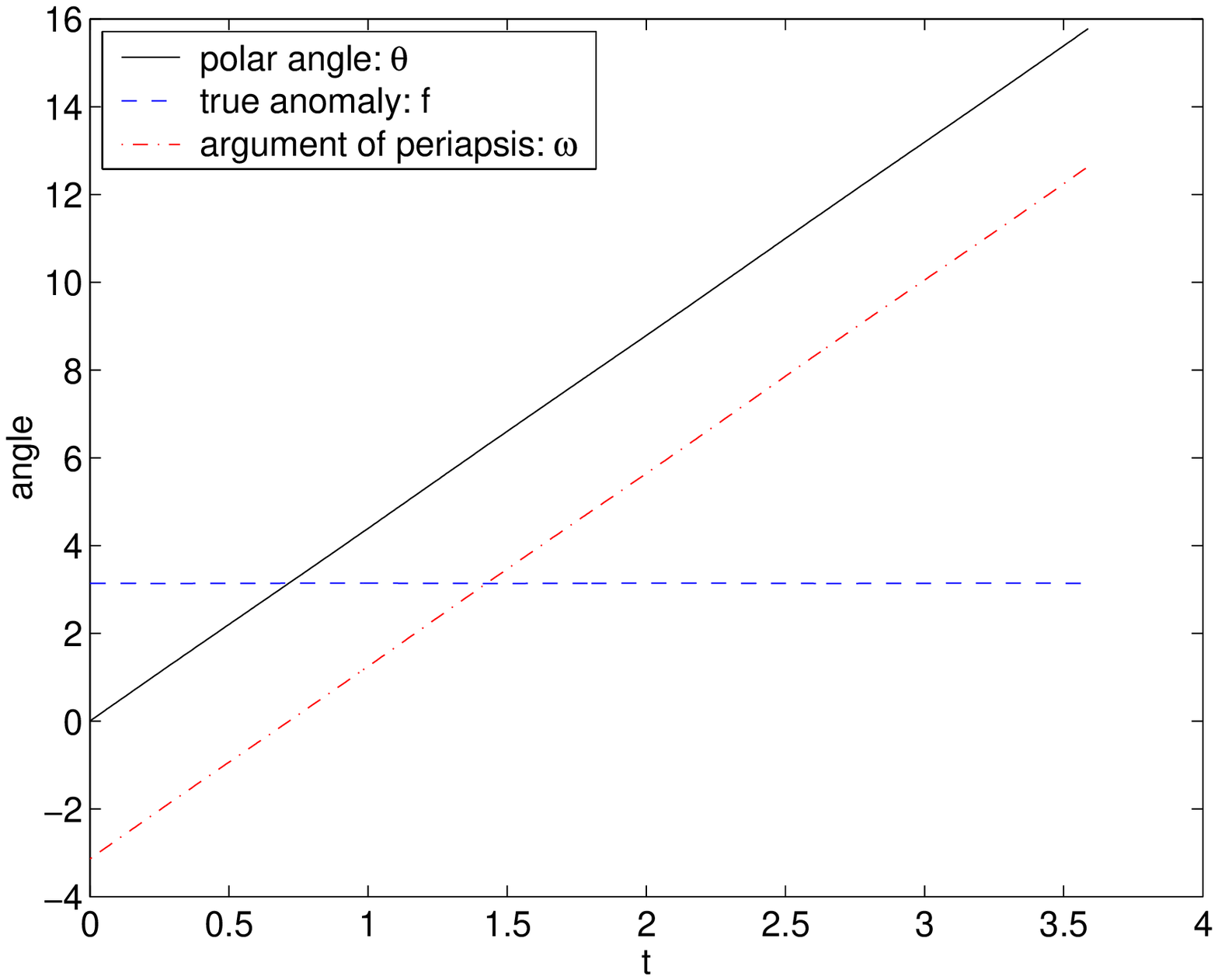}
\caption{$\omega(f)$ (left) and $\theta(t), \ \omega(t), \ f(t)$ (right) for $E = -0.9372$}
\label{ecchern33}
\end{center}
\end{figure}

\begin{figure}[ht!]
\begin{center}
\includegraphics[width=2.6in]{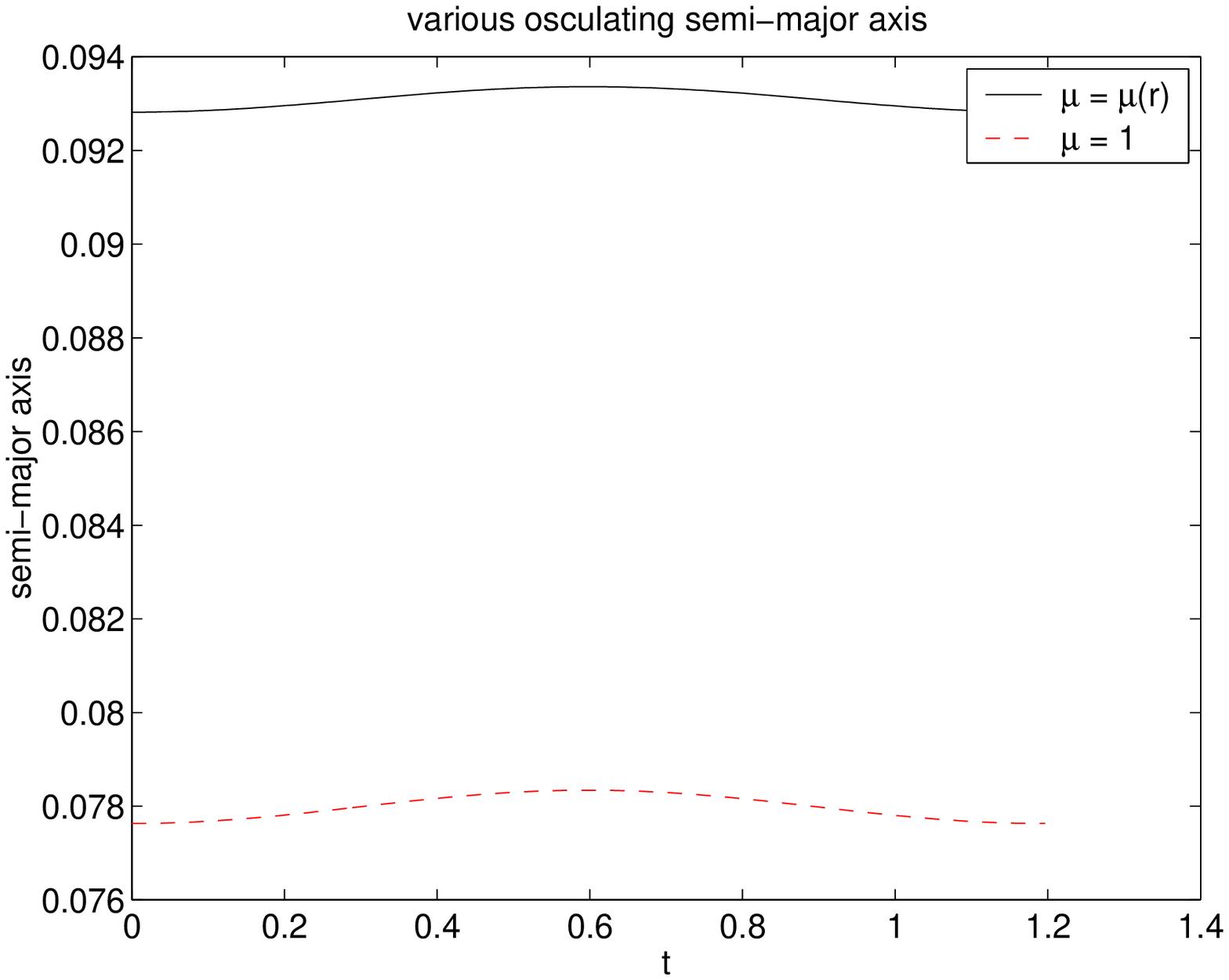}
\includegraphics[width=2.6in]{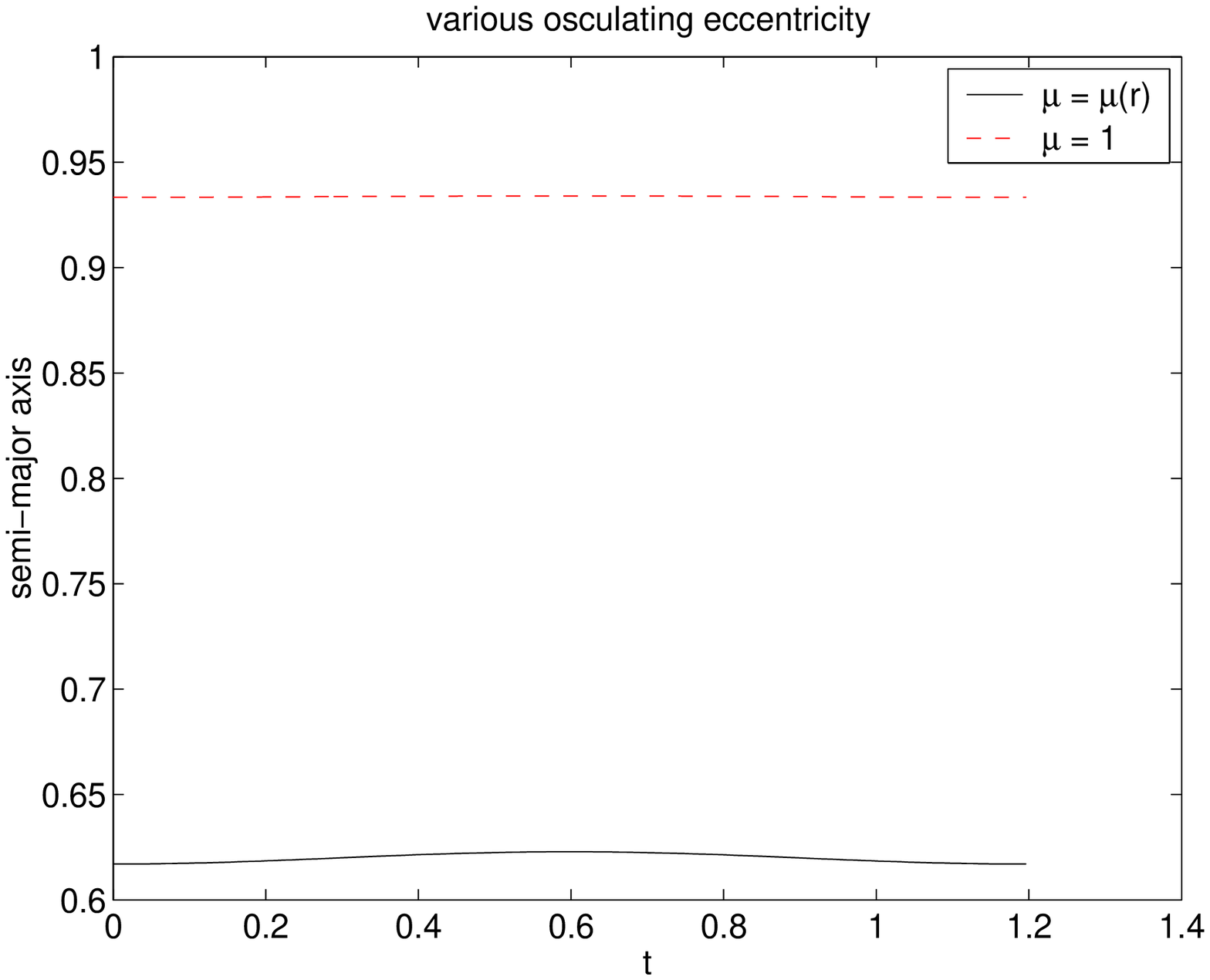}
\caption{Osculating orbital elements for $E = -0.9372$}
\label{ecchern35}
\end{center}
\end{figure}

Finally, when observing the semi-major axis and eccentricity of the orbit (Fig. \ref{ecchern35}), we see something counterintuitive.
The osculating eccentricity is close to $0.62$.  If one, on the other hand, used a classical definition of osculating orbital
elements, as previously discussed, the osculating eccentricity would be close to 0.94.  We thus see a circular orbit (in inertial space)
with high osculating eccentricity.  The osculating ellipse is a highly eccentric one that always touches the true path at apoapsis.  In this
way, the osculating ellipse rotates synchronously with the particle so that the particle is always at apoapsis
and the true motion is a circular path.

\section{Stability Analysis of Equatorial Rosettes in Axi-symmetric Potentials}

In this section we will consider the stability of the planar equatorial motion of a particle in an axi-symmetric
potential.  On the plane of symmetry, the potential reduces to a two-dimensional central force field.  
Since the motion is periodic with respect to the eccentric frame, the technique justifies application
of Floquet theory for the stability analysis of the in-plane motion, where the period used
is the period of the orbit in the eccentric frame.

\subsection{Floquet Theory}

Suppose now we are considering a conservative system with axi-symmetric equipotential contours, so that, in cylindrical coordinates,
the potential energy is given by $U(r,z)$.  We can define a variable gravitational parameter by $\mu(r,z) = r U(r,z)$.  The Hamiltonian of 
the system is thus given by:
\[
H = \frac 1 2 \left( v_r^2 + \frac{h_z^2}{r^2} + v_z^2 \right) - U(r,z) 
\]
where the symplectic coordinates are $\langle r, v_r, \theta, h_z, z, v_z \rangle$.  The equations of motion are:
\beas
\dot r = v_r &\qquad& \dot v_r = \frac{h_z^2}{r^3} + \parsh{U(r,z)}{r}  \\  
\dot \theta = \frac{h_z}{r^2} &\qquad & \dot h_z = 0 \\
\dot z = v_z & & \dot v_z = \parsh{U(r,z)}{z}.       
\eeas

In particular, equatorial motion with $z \equiv 0$ is well defined, and this reduces to the problem of motion in a central
force field.  The resulting motion follows a rosette-shaped path in the equatorial plane.  We now ask whether this motion is stable 
under a small out-of-plane perturbation $\delta z$.  To do this, we consider the $2 \times 2$ out-of-plane State Transition Matrix (STM) $\Phi$, which,
by definition, gives:
\[
 \left[ \begin{array}{c} \delta z(t) \\ \delta \dot z(t) \end{array} \right] = \Phi \cdot 
 \left[ \begin{array}{c} \delta z(0) \\ \delta \dot z(0) \end{array} \right].
\]
The STM is determined by integrating the following differential equation:
\[
\dot \Phi(t) = \left[ \begin{array}{cc} 0 & 1 \\ U_{zz}(r,0) & 0 \end{array} \right] \cdot \Phi(t), \qquad \Phi(0) = \left[ \begin{array}{cc} 1 & 0 \\ 0 & 1 
\end{array} \right],
\]
where the coefficient matrix is evaluated along the nominal orbit ($z = 0$) in the equatorial plane.  Since the equatorial
motion reduces to a central force field problem, there exists an eccentric frame decomposition, in which the motion is periodic.
Let $T$ be the period of a single orbit in the eccentric frame.  The coefficient matrix, above, only depends on $r(t)$, and thus
it is $T$-periodic.  We are therefore justified in using Floquet theory in the stability analysis.  Let $\la_1, \la_2$ be the 
eigenvalues of $\Phi(T)$.  Since $\la_1 \la_2 = 1$, either the eigenvalues are complex conjugates on the unit circle in the complex plane,
or real-valued with $\la_2 = \la_1^{-1}$.  The rosettal motion on the equatorial plane is therefore stable if and only if both
eigenvalues are on the unit circle.  A bifurcation from stable to unstable must occur when $\la_1 = \la_2 = 1$.

\subsection{Application to a toy axi-symmetric potential}

To show how this theory might be applied, we will consider the following toy potential:
\[
U(R)  =  \frac 1 R \left( 1 - \frac{1}{1 + R} \right) = \frac{\mu(R)}{R}, \ \  \mbox{where } \  \ R = \sqrt{r^2 + \frac{z^2}{a^2}}.
\]
Here, $r = \sqrt{x^2 + y^2}$ and $a>0$ is a parameter.  When $a < 1$, the potential is oblate spheroidal,
and when $a > 1$, it is prolate.  This potential is motivated
by replacing $r$ with $R$ in \eqref{ecc15}, but does not have direct physical significance.  Its utility to us
is only to illustrate the theory.  The potential reduces to the Hernquist potential when restricted
to the equatorial plane.  The question now arises, for various values of the parameter $a$, when is the
equatorial motion due to out of plane perturbations stable?  Clearly, for $a=1$, the motion is stable due to the
angular momentum integral.  To proceed, we compute the out of plane dynamics:
\[
\ddot z = \parsh{U}{z} = \frac{-\mu(R)}{R^3} \frac{z}{a^3} + \frac{\mu'(R)}{R} \frac{z}{Ra^2}.
\]
For a small perturbation $\delta z$, we obtain:
\[
\delta \ddot z = \omega(t) \delta z = \left( \frac{-\mu(r)}{a^2 r^3} + \frac{\mu'(r)}{a^2 r^2} \right) \delta z.
\]
The coefficient $\omega(t)$ is a function of time because we have an explicit solution for $r(t)$ for the 
nominal motion along the equatorial plane.  The out of plane State Transition Matrix (STM) for $\langle \delta z, 
\delta \dot z \rangle^T$ can then be written as:
\[
\dot \Phi = \left[ \begin{array}{cc}
0 & 1 \\
\omega(t) & 0
\end{array} \right] \cdot \Phi, \qquad \Phi(0) = \left[ \begin{array}{cc}
1 & 0 \\
0 & 1 \end{array} \right].
\]
The STM $\Phi$ can now be integrated along with the nominal planar solution, as its dynamic coefficient matrix
depends only on $r(t)$.  We can now systematically integrate $\dot \Phi$ between $t \in [0, T]$,
where $T$ is the eccentric frame orbital period.  Computing the eigenvalues of $\Phi(T)$ 
reveals the stability of the planar equatorial orbit.

\begin{figure}[h]
\begin{center}
\includegraphics[width=5in]{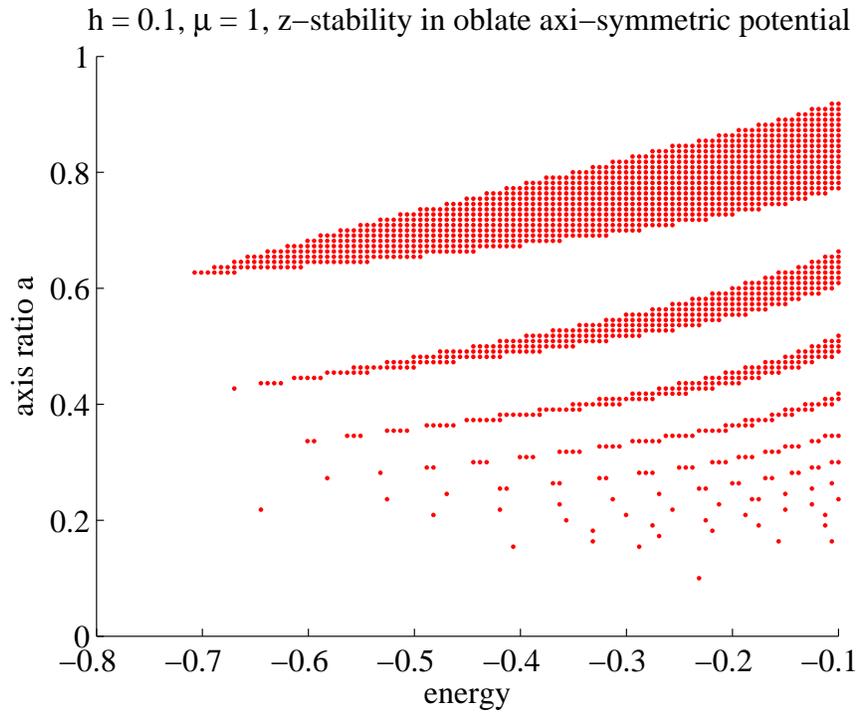}
\end{center}
\caption{Equatorial stability plot for oblate potential}
\label{eccfig14}
\end{figure}

\begin{figure}[h]
\begin{center}
\includegraphics[width=5in]{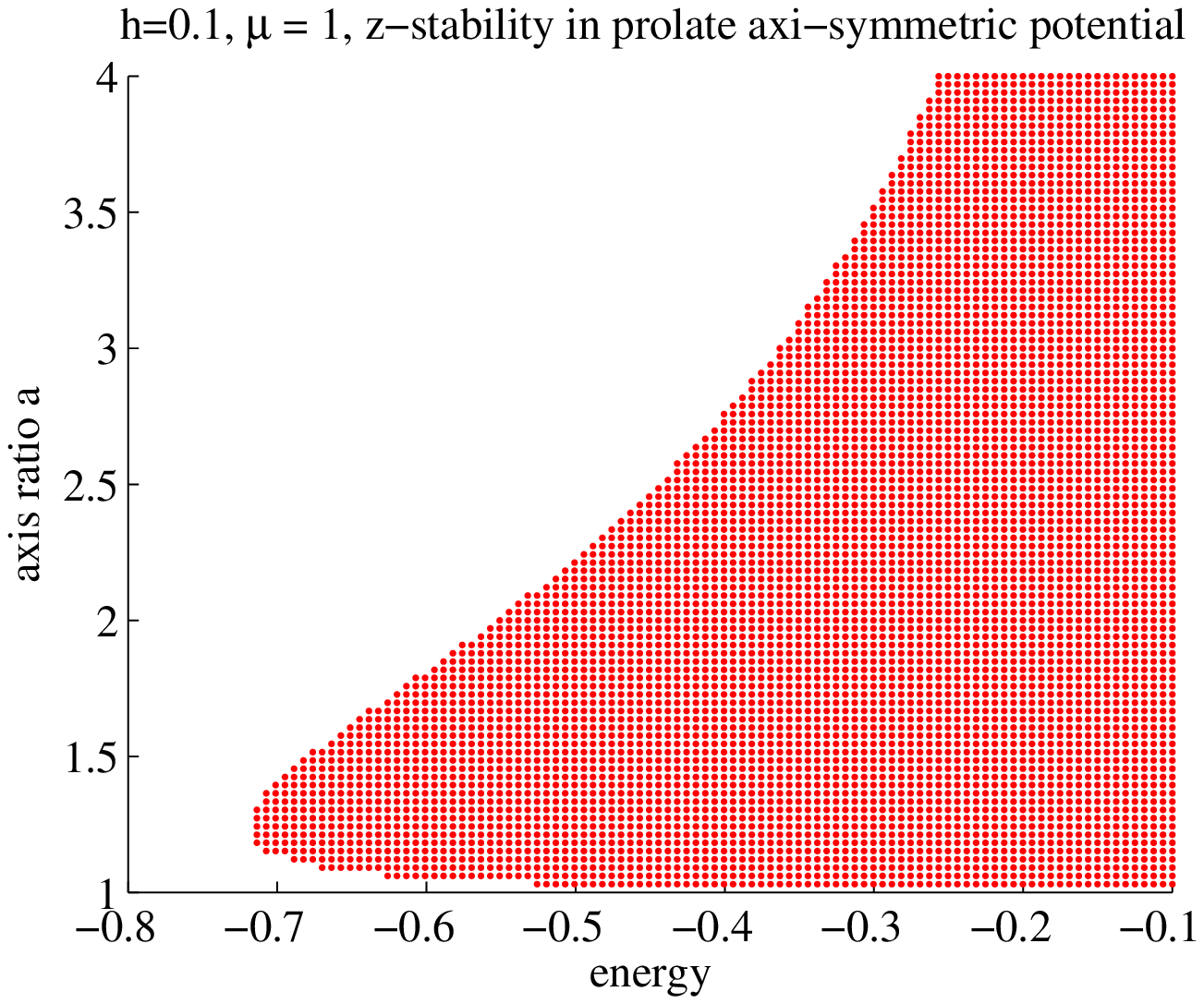}
\end{center}
\caption{Equatorial stability plot for prolate potential}
\label{eccfig15}
\end{figure}

We followed this procedure for a sampling of different energy levels and axis ratios $1:1:a$.
The result for oblate potentials is shown in Fig. \ref{eccfig14}.  The grid points with dots
correspond to parameter values of $a$ and $E$ for which the equatorial motion is unstable.  
For prolate potentials, the result is similarly depicted in Fig. \ref{eccfig15}.  Notice in both plots,
for $a=1$, that the planar motion is stable for all energies.  

\section{Conclusion}

In this paper we presented a preferred, nonuniformly rotating frame that exists for motion
in any central force field, with respect to which the orbital motion is periodic.  We showed that 
for high values of energy, the particle trajectories in the eccentric frame make circulations.  However,
when the energy drops beneath a certain critical level, the trajectories follow librations
in the eccentric frame.  This is not a true bifurcation of the system, as there is no distinguishable
physical change when the orbits are viewed with respect to inertial space, yet it is a necessary
transition that must occur as one nears the minimum circular orbit energy.  For circular orbits (in 
inertial space), motion in the eccentric frame degenerates from librations to a single fixed point.
For this case, the eccentric frame rotates at a constant rate, and a circular orbit
in inertial space is observed.  Further we showed that, even in the case of a circular orbit,
the osculating eccentricity can be very high.  This occurs because the particle is ``stuck'' to 
the periapsis or apoapsis of the osculating ellipse.  The osculating ellipse has static (high) eccentricity
and rotates at a uniform rate.  We also presented a model for the potential energy
of a Hernquist galaxy with a central black hole, analyzed the rosette-shaped orbits,
and then compared them to the orbit as seen from the eccentric frame for various parameter values.
Finally we indicated how one might use the eccentric frame to determine stability of planar
orbits in axi-symmetric potentials.

\section*{References}


\noindent
Adams, F.C., Bloch, A.M.: Orbits in extended mass distributions:
general results and the spirographic approximation.  The 
Astrophysical Journal. 629, 204-218 (2005)

\noindent
Arnold, V.I.: Mathematical Aspects of Classical and 
Celestial Mechanics. Springer-Verlag (1989)

\noindent
Arnold, V.I.: Dynamical Systems III: Mathematical Aspects of
Classical and Celestial Mechanics. Springer-Verlag (1993)

\noindent
Brouwer, D., Clemence, G.M.: 
Methods of Celestial Mechanics. Academic (1961)

\noindent
Craig, S., Diacu, F., Lacomba, E. A., Perez, E.: On the Anisotropic
Manev Problem.  Journal of Mathematical Physics, 40, 1359 - 1375 (1999)

\noindent
Diacu, F., Santoprete, M.: Nonintegrability and chaos in the
anisotropic Manev problem.  Physica D, 156, 39 - 52 (2001)

\noindent
Greenwood, D.T.: 
Classical Dynamics, Dover (1997)

\noindent
Hernquist, L.: 
An analytical model for spherical galaxies and bulges.
The Astrophysical Journal. 356, 359-364 (1990)

\noindent
Poon, M.Y., Merritt, D.: 
A self-consistent study of triaxial
black hole nuclei. The Astrophysical Journal. 606, 774-787 (2004)

\noindent
Roy, A.E.:
Orbital Motion. Institute of Physics (1988)

\noindent
Whittaker, E.T.: 
A Treatise on the Analytical Dynamics
of Particles and Rigid Bodies. Cambridge University Press (1988)


\end{document}